\documentclass[aps,prd,a4paper,twocolumn,amsmath,showpacs,superscriptaddress,nofootinbib,preprintnumbers]{revtex4-1}

\usepackage{verbatim}
\usepackage{epsfig}
\usepackage{graphicx}
\usepackage{booktabs}
\usepackage{multirow}
\usepackage{dcolumn}
\usepackage{amsmath}
\usepackage{mathtools}
\usepackage{amsfonts}
\usepackage{amssymb}
\usepackage{epstopdf}
\usepackage{bm}
\usepackage{siunitx}
\usepackage{braket}
\usepackage{enumitem}
\usepackage{soul}
\usepackage[table]{xcolor}
\definecolor{navyblue}{rgb}{0.0, 0.0, 0.5}
\definecolor{royalblue}{rgb}{0.25, 0.41, 0.88}
\definecolor{cadmiumgreen}{rgb}{0.0, 0.42, 0.24}
\definecolor{blue-violet}{rgb}{0.54, 0.17, 0.89}
\definecolor{darkviolet}{rgb}{0.58, 0.0, 0.83}
\definecolor{orange(colorwheel)}{rgb}{1.0, 0.5, 0.0}

\usepackage{hyperref}
\hypersetup{colorlinks=true,linkcolor=royalblue,citecolor=magenta}
\usepackage{pifont}
\usepackage{dcolumn}
\usepackage{colortbl}

\newcommand\ee{\end{equation}}
\newcommand\be{\begin{equation}}
\newcommand\eea{\end{eqnarray}}
\newcommand\bea{\begin{eqnarray}}

\renewcommand\({\left(}
\renewcommand\){\right)}

\newcommand{\ns}{n_{\rm s}}

\newcommand\limit[1]{#1\%\,\mathrm{CL}}

\newcommand{\tu}{\textup}
\newcommand{\dif}{\mathrm{d}}
\newcommand\ie{{\it i.e.}~}

\newcommand\eq[1]{Eq.~\eqref{eq:#1}}
\newcommand\sect[1]{Sec.~\ref{sec:#1}}
\newcommand\fig[1]{Fig.~\ref{fig:#1}}
\newcommand\tab[1]{Tab.~\ref{tab:#1}}

\renewcommand{\vec}{\bm}

\newcommand{\alphas}{\alpha_\mathrm{s}}
\newcommand{\betas}{\beta_\mathrm{s}}
\newcommand{\lnrat}{\log\frac{k}{k_\star}}

\newcommand\vertsp{\rule[-2mm]{1mm}{0mm} &}
\newcommand\horsp{\rule[-1.5mm]{0mm}{4.125mm}}
\newcommand\morehorsp{\rule[-2.25mm]{0mm}{6mm}}

\newcommand{\siround}[2]{\num[round-mode=places,group-digits=false,round-precision=#2]{#1}}

\definecolor{magenta(process)}{rgb}{1.0, 0.0, 0.56}

\definecolor{darkspringgreen}{rgb}{0.09, 0.45, 0.27}

\definecolor{royalblue(web)}{rgb}{0.25, 0.41, 0.88}
\begin{document}

\title{Running the running}

\author{Giovanni Cabass}
\affiliation{Physics Department and INFN, Universit\`a di Roma 
	``La Sapienza'', P.le\ Aldo Moro 2, 00185, Rome, Italy}

\author{Eleonora Di Valentino}
\affiliation{Institut d’Astrophysique de Paris (UMR7095: CNRS \& UPMC-Sorbonne Universities), F-75014, Paris, France}
	
\author{Alessandro Melchiorri}
\affiliation{Physics Department and INFN, Universit\`a di Roma 
	``La Sapienza'', P.le\ Aldo Moro 2, 00185, Rome, Italy}
	
\author{Enrico Pajer}
\affiliation{Institute for Theoretical Physics and Center for Extreme Matter and Emergent Phenomena,
	Utrecht University, Princetonplein 5, 3584 CC Utrecht, The Netherlands}

\author{Joseph Silk}
\affiliation{Institut d’Astrophysique de Paris (UMR7095: CNRS \& UPMC-Sorbonne Universities), F-75014, Paris, France}
\affiliation{AIM-Paris-Saclay, CEA/DSM/IRFU, CNRS, Univ. Paris VII, F-91191 Gif-sur-Yvette, France}
\affiliation{Department of Physics and Astronomy, The Johns Hopkins University Homewood Campus, Baltimore, MD 21218, USA }
\affiliation{BIPAC, Department of Physics, University of Oxford, Keble Road, Oxford OX1 3RH, UK}

\begin{abstract}
\noindent We use the recent observations of Cosmic Microwave Background temperature and polarization anisotropies provided by the \emph{Planck} satellite experiment to place constraints on the running $\alpha_\mathrm{s} = \mathrm{d}n_{\mathrm{s}} / \mathrm{d}\log k$ and the running of the running $\beta_{\mathrm{s}} = \mathrm{d}\alpha_{\mathrm{s}} / \mathrm{d}\log k$ of the spectral index $n_{\mathrm{s}}$ of primordial scalar fluctuations. We find $\alpha_\mathrm{s}=0.011\pm0.010$ and $\beta_\mathrm{s}=0.027\pm0.013$ at $68\%\,\mathrm{CL}$, suggesting the presence of a running of the running at the level of two standard deviations. We find no significant correlation between $\beta_{\mathrm{s}}$ and foregrounds parameters, with the exception of the point sources amplitude at $143\,\mathrm{GHz}$, $A^{PS}_{143}$, which shifts by half sigma when the running of the running is considered. 
We further study the cosmological implications of such preference for $\alpha_\mathrm{s},\beta_\mathrm{s}\sim 0.01$ by including in the analysis the lensing amplitude $A_L$, the curvature parameter $\Omega_k$, and the sum of neutrino masses $\sum m_{\nu}$. We find that when the running of the running is considered, %the 
\emph{Planck} data are more compatible with the standard expectations of $A_L = 1$ and $\Omega_k = 0$ but still %hinting 
hint at possible deviations. The indication for $\beta_\mathrm{s} > 0$ survives at two standard deviations when external datasets such as BAO and CFHTLenS are included in the analysis, and persists at $\sim 1.7$ standard deviations when CMB lensing is considered. We discuss the possibility of constraining $\beta_\mathrm{s}$ with current and future measurements of CMB spectral distortions, showing that an experiment like PIXIE could provide strong constraints on $\alpha_\mathrm{s}$ and $\beta_\mathrm{s}$.
\end{abstract}


\maketitle

\twocolumngrid

%%%%%%%%%%%%%%%%%%%%%%%%%%%%%%%%%%%%%%%%%%%%%%%%%%%%%%%%%%%%%%%%%%%%%%%%%%%%%%%%%

\section{Introduction}
\label{sec:introduction}

\noindent The recent measurement of the Cosmic Microwave Background (CMB) anisotropies provided by the \emph{Planck} satellite mission (see \cite{Ade:2015xua, Ade:2015lrj}, for example) have provided a wonderful confirmation of the standard $\Lambda$CDM cosmological model. However, when the base model is extended and other cosmological parameters are let free to vary, a few ``anomalies'' are present in the parameter %s 
values that, %albeit with a significance at about only two standard deviations, 
even if their significance is only at the level of two standard deviations, deserve further investigation.

First of all, the parameter $A_{L}$, that measures the amplitude of the lensing signal
in the CMB angular spectra \cite{Calabrese:2008rt}, has been found larger than the standard value with
$A_{L}=1.22\pm0.10$ at $\limit{68}$ ($A_{L}=1$ being the expected value in $\Lambda$CDM) 
from \emph{Planck} temperature and polarization angular spectra \cite{Ade:2015xua}. A value of $A_{L}$ larger than one is difficult
to accommodate in $\Lambda$CDM, and several solutions have been proposed
as modified gravity \cite{edmg,Huang:2015srv}, neutrino anisotropies \cite{Gerbino:2013ova}, 
and compensated isocurvature perturbations \cite{cip}. 
Combining \emph{Planck} with data from the Atacama Cosmology Telescope (ACT) and the South Pole Telescope (SPT) 
to better constrain the foregrounds, Couchot et al. \cite{couchot}, found a consistency with $A_L=1$. However
the compatibility of the CMB datasets used is unclear. 
More recently Addison et al. \cite{addison} have found that including the $A_{L}$ parameter solves the tension
between \emph{Planck} and WMAP9 on the value of the derived cosmological parameters.

As shown in \cite{Ade:2015xua}, the $A_{L}$ anomaly persists when the \emph{Planck} data is combined with Baryonic Acoustic Oscillation surveys (BAO), it is enhanced when the CFHTLenS {shear lensing survey} is included, but it practically disappears when
CMB lensing from \emph{Planck} trispectrum observations are considered.
The $A_{L}$ anomaly is also still present in a $12$-parameter extended $\Lambda$CDM analysis of the 
\emph{Planck} dataset (see \cite{ems}), showing no significant correlation with extra parameters such as the dark energy equation of state $w$, 
the neutrino mass, and the neutrino effective number $N_\mathrm{eff}$.

Second, the \emph{Planck} dataset prefers a positively curved universe, again at about 
two standard deviations with $\Omega_k = -0.040\pm0.020$ at $\limit{68}$.
This ``anomaly'' is not due to an increased parameter volume effect but, as stated in
\cite{Ade:2015lrj}, curvature provides a genuine better fit to the data with an improved
fit of $\Delta \chi^2 \sim 6$. When BAO data is included, however, {the curvature of the universe} is
again compatible with zero with the stringent constraint 
$\Omega_k=-0.000\pm0.005$ at $\limit{95}$.

The fact that both the $A_{L}$ and $\Omega_k$ anomalies disappear when
reliable external datasets are included suggests that their origin might be 
a systematic or that they are produced by a different physical effect
than lensing or curvature.

In this respect it is interesting to note that a third parameter is constrained to anomalous values from the \emph{Planck} data. 
The primordial scalar spectral index $\ns$ of scalar perturbations is often assumed to be independent of scale. However, since 
some small scale-dependence is expected,\footnote{\emph{E.g.}, we expect a running of the tilt $\ns$ of order $(1-\ns)^2$ in slow-roll inflation.} %it can be expanded as 
we can expand the dimensionless scalar power spectrum $\Delta^2_\zeta(k) = k^3P_\zeta(k)/2\pi^2$ as 
\begin{equation}
\label{eq:Delta_of_k}
\Delta^2_\zeta(k) = A_\mathrm{s}\(\frac{k}{k_\star}\)^{\ns-1 + \frac{\alphas}{2}\lnrat + \frac{\betas}{6}\(\lnrat\)^2}\,\,,
\end{equation}
where $\alphas$ is the running of the spectral index, $\betas$ is the running of the running, and $k_\star = 0.05\,\mathrm{Mpc}^{-1}$. 

The \emph{Planck} temperature and polarization data analysis presented in
\cite{Ade:2015lrj}, while providing a small indication for a {\it positive} running different from zero ($\alphas=0.009\pm0.010$ at $\limit{68}$), suggests also the presence of a running of the running at the level of two standard deviations 
($\betas=0.025\pm0.013$ at $\limit{68}$). The inclusion of a running of the running improves the fit to the \emph{Planck}
temperature and polarization data by $\Delta\chi^2\sim 5$ \textcolor{black}{with respect to the $\Lambda$CDM model. Therefore we do not expect that such anomaly is due to the increased parameter volume, and could be a hint of possible 
new physics beyond the standard model}. A discussion of the impact of this anomaly on inflationary models
has been presented in \cite{menabetas, Huang:2015cke}.

Given this result, it is %definitely 
timely to discuss the possible
correlations between these three anomalies, $\betas$, $A_{L}$ and $\Omega_k$ and see, for example, if
one of them vanishes if a second one is considered at the same time in the analysis.
Moreover \textcolor{black}{(related to the above points), it is necessary to investigate in more detail how the inclusion of $\betas$ helps giving a better fit to the data, and} test if the indication for the running of the running
survives when additional datasets as BAO or lensing (CMB and shear) 
are considered. This is the goal of this paper. 

We structure the discussion as follows. In the next section we will describe the analysis method and
the cosmological datasets used. In \sect{results} we present our results
and discuss possible correlations between $\betas$, $A_{L}$
and $\Omega_k$. We also investigate the possibility that a running of the running affects current 
and future measurements of CMB spectral distortions, comparing our results with those of \cite{Powell:2012xz}. Finally, in \sect{conclusions} we derive our conclusions.

\section{Method}
\label{sec:method}

\noindent We perform a Monte Carlo Markov Chain (MCMC) analysis of the most recent cosmological 
datasets using the publicly available code \texttt{cosmomc}~\cite{Lewis:2002ah, Lewis:2013hha}. 
We consider the $6$ parameters of the standard $\Lambda$CDM model, \emph{i.e.} 
the baryon $\omega_\mathrm{b}\equiv\Omega_\mathrm{b} h^2$ and cold dark matter $\omega_\mathrm{c}\equiv\Omega_\mathrm{c} h^2$ energy densities,
the angular size of the horizon at the last scattering surface $\theta_\mathrm{MC}$, the 
optical depth $\tau$, the amplitude of primordial scalar perturbations $\log (10^{10}A_\mathrm{s})$ and the scalar spectral index $\ns$. 
We extend this scenario by including the running of the scalar spectral
index $\alphas$ and the running of the running $\betas$. We fix the
pivot scale at $k_\star=0.05\,\mathrm{Mpc}^{-1}$. This is our baseline cosmological model, that we will call ``base'' in the following. 
Moreover, as discussed in the introduction, we also consider separate variation in the
lensing amplitude $A_{L}$, in the curvature density $\Omega_k$ and in the sum of neutrino masses $\sum m_\nu$.

The main dataset we consider, to which we refer as ``\emph{Planck}'', is based on CMB temperature and polarization anisotropies. We analyze the temperature and polarization \emph{Planck} likelihood \cite{Aghanim:2015xee}: more precisely, we make use of the $TT$, $TE$, $EE$ high-$\ell$ likelihood together with the $TEB$ pixel-based low-$\ell$ likelihood. The additional datasets we consider are the following:
\begin{itemize}[leftmargin=*]
\item {\emph{Planck} measurements of the lensing potential power spectrum $C^{\phi\phi}_\ell$ \cite{Ade:2015zua};}
\item weak gravitational lensing data of the CFHTLenS survey \cite{Heymans:2012gg, Erben:2012zw}, taking only wavenumbers %with 
$k\leq 1.5 h\,\mathrm{Mpc}^{-1}$\cite{Ade:2015xua, Kitching:2014dtq};
\item {Baryon Acoustic Oscillations (BAO): the surveys included are 6dFGS \cite{Beutler:2011hx}, SDSS-MGS \cite{Ross:2014qpa}, BOSS LOWZ \cite{Anderson:2013zyy} and CMASS-DR11 \cite{Anderson:2013zyy}. This dataset will help to break geometrical degeneracies when we let $\Omega_k$ free to vary.}
\end{itemize}

%\begin{figure*}
%\begin{center}
%\begin{tabular}{c}
%\includegraphics[width=0.75\textwidth]{tt_spectra_bf-with_errors-check-final.pdf} %\\
%\includegraphics[width=0.75\textwidth]{tt_spectra_residuals_bf-with_errors-check-final.pdf}
%\end{tabular}
%\end{center}
%\caption{\footnotesize{CMB temperature power spectra as measured by \emph{Planck}, with the best-fit from the $\Lambda\mathrm{CDM}$, $\Lambda\mathrm{CDM} + \alphas + \betas$, $\Lambda\mathrm{CDM} + A_L$ and $\Lambda\mathrm{CDM} + \Omega_k$ models. We see how the preference for large positive values of $\alphas$ and $\betas$ (that lower the power at large scales), with respect to the $\Lambda\mathrm{CDM}$ model, is driven by the low $TT$ quadrupole. We refer to the main text for a discussion.}}
%\label{fig:bestfits}
%\end{figure*}

\begin{table*}[!hbtp]
%\footnotesize
\begin{center}
\begin{tabular}{lcccc}
\toprule
\horsp
$\mathrm{base}$ \vertsp \emph{Planck} \vertsp + lensing \vertsp + WL \vertsp + BAO \\
\hline
\morehorsp
$\Omega_\mathrm{b}h^2$ \vertsp ${\siround{0.02216}{5}}\pm{\siround{0.00017}{5}}$ \vertsp ${\siround{0.02215}{5}}\pm{\siround{0.00017}{5}}$ \vertsp ${\siround{0.02221}{5}}\pm{\siround{0.00017}{5}}$ \vertsp ${\siround{0.02224}{5}}\pm{\siround{0.00015}{5}}$ \\
\morehorsp
$\Omega_\mathrm{c}h^2$ \vertsp ${\siround{0.1207}{4}}\pm{\siround{0.0015}{4}}$ \vertsp ${\siround{0.1199}{4}}\pm{\siround{0.0015}{4}}$ \vertsp ${\siround{0.1197}{4}}\pm{\siround{0.0014}{4}}$ \vertsp ${\siround{0.1196}{4}}\pm{\siround{0.0011}{4}}$ \\
\morehorsp
$100\theta_\mathrm{MC}$ \vertsp ${\siround{1.0407}{5}}\pm{\siround{0.00032}{5}}$ \vertsp ${\siround{1.0408}{5}}\pm{\siround{0.00032}{5}}$ \vertsp ${\siround{1.04078}{5}}\pm{\siround{0.00032}{5}}$ \vertsp ${\siround{1.04082}{5}}^{+\siround{0.00029}{5}}_{-\siround{0.0003}{5}}$ \\
\morehorsp
$\tau$ \vertsp ${\siround{0.091}{3}}\pm{\siround{0.019}{3}}$ \vertsp ${\siround{0.064}{3}}\pm{\siround{0.014}{3}}$ \vertsp ${\siround{0.086}{3}}\pm{\siround{0.019}{3}}$ \vertsp ${\siround{0.096}{3}}\pm{\siround{0.018}{3}}$ \\
\morehorsp
$H_0$ \vertsp ${\siround{66.88}{2}}\pm{\siround{0.68}{2}}$ \vertsp ${\siround{67.16}{2}}\pm{\siround{0.67}{2}}$ \vertsp ${\siround{67.29}{2}}^{+\siround{0.66}{2}}_{-\siround{0.65}{2}}$ \vertsp ${\siround{67.36}{2}}^{+\siround{0.49}{2}}_{-\siround{0.48}{2}}$ \\
\morehorsp
$\log(10^{10} A_\mathrm{s})$ \vertsp ${\siround{3.118}{3}}\pm{\siround{0.037}{3}}$ \vertsp ${\siround{3.061}{3}}\pm{\siround{0.026}{3}}$ \vertsp ${\siround{3.104}{3}}^{+\siround{0.038}{3}}_{-\siround{0.037}{3}}$ \vertsp ${\siround{3.125}{3}}\pm{\siround{0.036}{3}}$ \\
\morehorsp
$n_\mathrm{s}$ \vertsp ${\siround{0.9582}{4}}^{+\siround{0.0055}{4}}_{-\siround{0.0054}{4}}$ \vertsp ${\siround{0.9607}{4}}\pm{\siround{0.0054}{4}}$ \vertsp ${\siround{0.9608}{4}}\pm{\siround{0.0055}{4}}$ \vertsp ${\siround{0.9613}{4}}^{+\siround{0.0046}{4}}_{-\siround{0.0047}{4}}$ \\
\morehorsp
$\alpha_\mathrm{s}$ \vertsp ${\siround{0.011}{3}}\pm{\siround{0.01}{3}}$ \vertsp ${\siround{0.012}{3}}\pm{\siround{0.01}{3}}$ \vertsp ${\siround{0.012}{3}}\pm{\siround{0.01}{3}}$ \vertsp ${\siround{0.01}{3}}\pm{\siround{0.01}{3}}$ \\
\morehorsp
$\beta_\mathrm{s}$ \vertsp ${\siround{0.027}{3}}\pm{\siround{0.013}{3}}$ \vertsp ${\siround{0.022}{3}}\pm{\siround{0.013}{3}}$ \vertsp ${\siround{0.026}{3}}\pm{\siround{0.013}{3}}$ \vertsp ${\siround{0.025}{3}}\pm{\siround{0.013}{3}}$ \\
\hline
\bottomrule
%\botrule
\end{tabular}
\caption{\footnotesize{$\limit{68}$ bounds %and $\limit{95}$ upper limits 
on $\Omega_\mathrm{b}h^2$, $\Omega_\mathrm{c}h^2$, $100\theta_\mathrm{MC}$, $\tau$, $H_0$, $\log(10^{10} A_\mathrm{s})$, $n_\mathrm{s}$, $\alpha_\mathrm{s}$, $\beta_\mathrm{s}$, for the listed datasets: the model is $\Lambda\mathrm{CDM} + \alpha_\mathrm{s} + \beta_\mathrm{s}$, $k_\star = 0.05\,\mathrm{Mpc}^{-1}$.}}
\label{tab:results}
\end{center}
\end{table*}

\section{Results}
\label{sec:results}

\noindent In Tab.~\ref{tab:results} we present the constraints on $\ns$, $\alphas$ and $\betas$
from the \emph{Planck} 2015 temperature and polarization data and in combination 
with BAO, cosmic shear and CMB lensing. 
As we can see, the Planck alone dataset provides an indication for $\betas>0$ at more
than two standard deviations with $\betas=0.027\pm0.013$ at $\limit{68}$. 

\begin{figure*}
\begin{center}
\begin{tabular}{c c}
\includegraphics[width=\columnwidth]{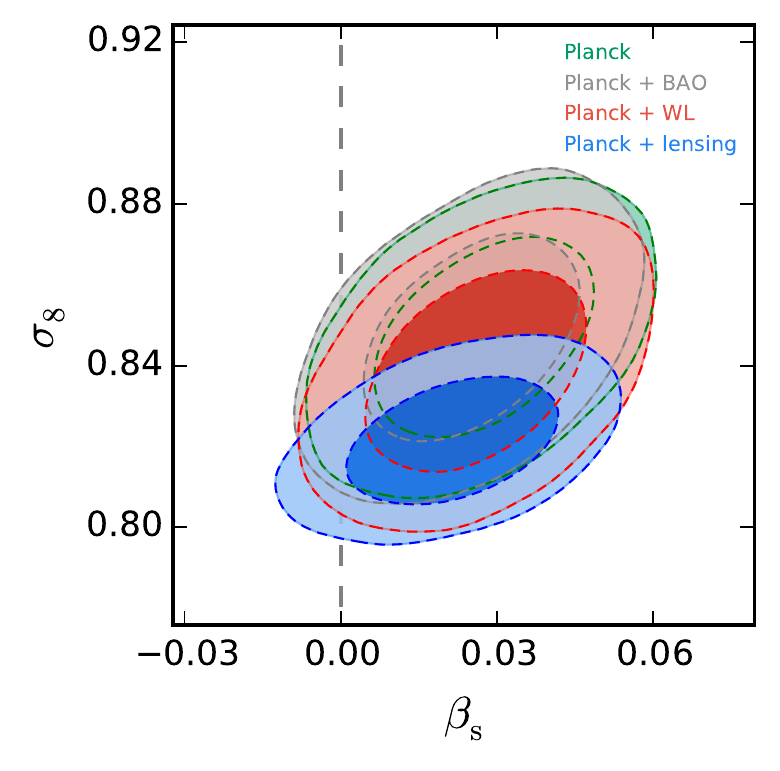}
&\includegraphics[width=\columnwidth]{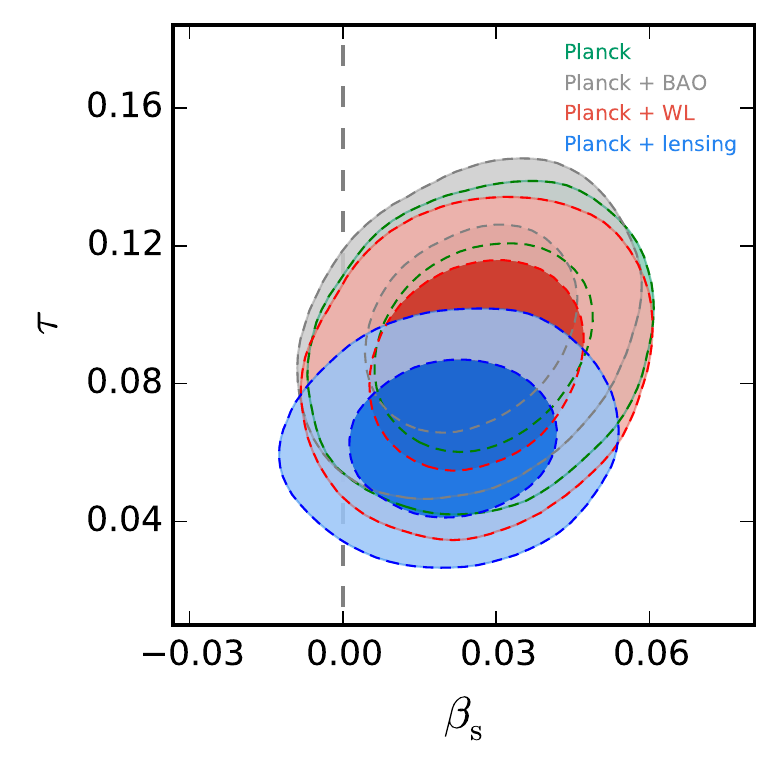}
\end{tabular}
\end{center}
\caption{\footnotesize{Constraints at $\limit{68}$ and $\limit{95}$ in the $\text{$\betas$ -- $\sigma_8$}$ plane (left panel)
and in the $\text{$\betas$ -- $\tau$}$ plane (right panel).}}
\label{fig:tau_sigma8-v-nrunrun}
\end{figure*}

It is interesting to investigate the impact of the inclusion of $\alphas$ and 
$\betas$ on the remaining $6$ parameters of the $\Lambda$CDM model.
Comparing our results with those reported in Table $3$ of \cite{Ade:2015lrj}, 
we see that there are no major shifts on the parameters. The
largest shifts are present for the scalar spectral index $\ns$, that is $\sim 0.9$ 
standard deviations {\it lower} when $\betas$ is included, and for
the reionization optical depth $\tau$ that is $\sim 0.9$ 
standard deviations {\it higher} with respect to the standard $\Lambda$CDM
scenario. A similar shift is also present for the value of the root mean square
density fluctuations on scales of $8 h\,\mathrm{Mpc}^{-1}$ (the $\sigma_8$ derived parameter), 
which is higher by about one standard deviation when $\betas$ is considered. 
In Fig.~\ref{fig:tau_sigma8-v-nrunrun} we plot the probability contour at $\limit{68}$ and
$\limit{95}$ for the several combinations of datasets in the $\text{$\betas$ -- $\sigma_8$}$ and $\text{$\betas$ -- $\tau$}$ planes respectively. Clearly, a new determination of $\tau$ from future large-scale polarization data as those expected from
the Planck HFI experiment could have an impact on the value of $\betas$.
On the other hand, this one sigma shift in $\tau$ with respect to $\Lambda$CDM shows that a 
large-scale measurement of CMB polarization does not fully provide a direct determination 
of $\tau$ but that some model dependence is present.

\begin{figure*}
\begin{center}
\begin{tabular}{c c}
\includegraphics[width=\columnwidth]{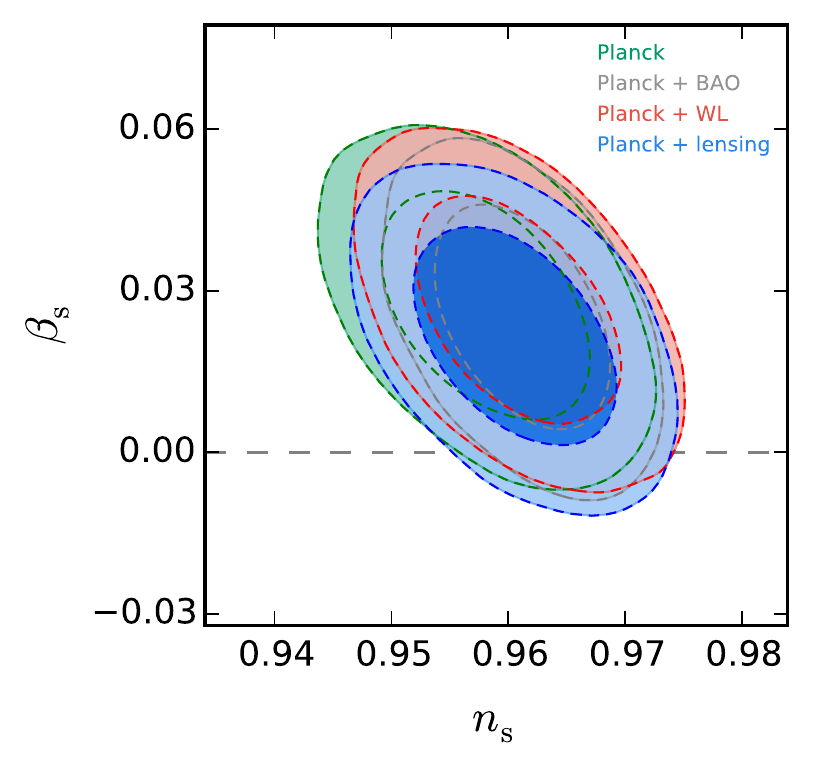}
&\includegraphics[width=\columnwidth]{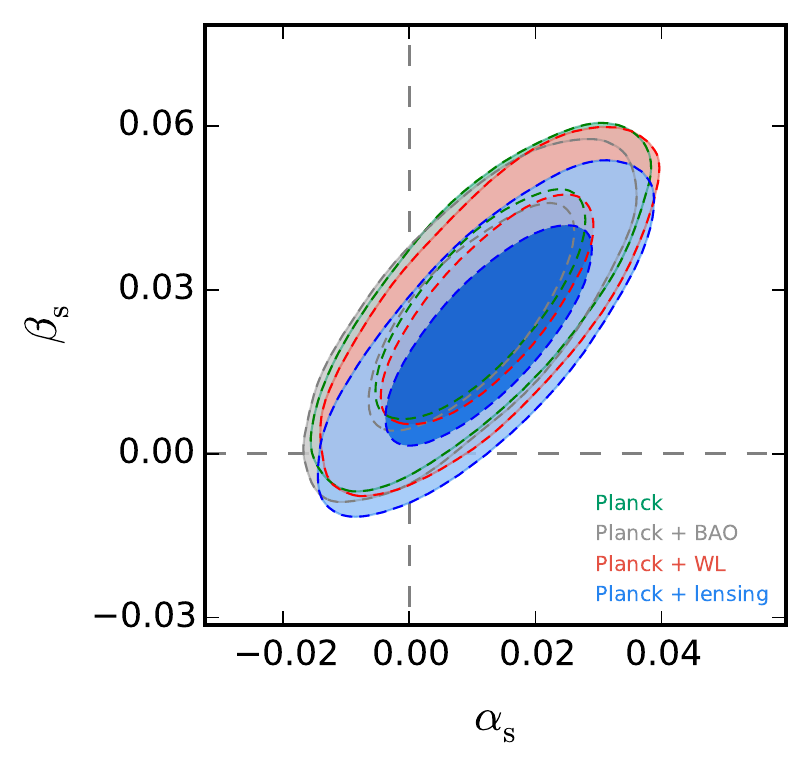}
\end{tabular}
\end{center}
\caption{\footnotesize{Likelihood constraints 
in the $\text{$\ns$ -- $\betas$}$ (left panel) and $\text{$\alphas$ -- $\betas$}$ (right panel) planes for
different combination of datasets, as discussed in the text.}}
\label{fig:ns_nrun-v-nrunrun}
\end{figure*}

Moreover, as expected, there is a strong correlation between $\alphas$ and
$\betas$. Because of this correlation, the running $\alphas$ is constrained to be positive, 
with $\alphas>0$ at more than $\limit{68}$ when $\betas$ is considered. 
This is a $\sim 1.3$ standard deviations shift on $\alphas$ if we compare this result
with the value obtained using the same dataset but fixing $\betas=0$ in Table 
$5$ of \cite{Ade:2015lrj}.
In Fig.~\ref{fig:ns_nrun-v-nrunrun} we plot the two dimensional likelihood constraints
in the $\text{$\ns$ -- $\betas$}$ and $\text{$\alphas$ -- $\betas$}$ planes respectively. 
As we can see, a correlation between the parameters is clearly present. \textcolor{black}{However, when $\alphas$ and possibly higher derivatives of the scalar tilt are left free to vary, the constraints %on these parameters 
will depend on the choice of the pivot scale $k_\star$ \cite{Cortes:2007ak}. We have therefore considered two additional values of $k_\star$, \emph{i.e.} $k_\star = 0.01\,\mathrm{Mpc}^{-1}$ and $k_\star = 0.002\,\mathrm{Mpc}^{-1}$: the resulting plots are shown in \sect{dependence_pivot} (where we also present a simple argument to explain the stability of $\sigma_{\betas}$ under change of $k_\star$), while \tab{pivots} shows the $\limit{68}$ constraints on $\ns$, $\alphas$ and $\betas$ (``base'' model, \emph{Planck} $TT$, $TE$, $EE$ + lowP dataset\footnote{A study of the impact of $k_\star$ when also $A_L$, $\sum m_\nu$ and $\Omega_K$ are varied is left to future work.}). From \tab{pivots} we see that, while the $1\sigma$ indication for $\alphas > 0$ disappears if we change $k_\star$ (becoming a $\sim2\sigma$ evidence for negative running), $\betas$ remains larger than $0$ at $\sim 2\sigma$.\footnote{We also note a $\sim1\sigma$ indication of blue tilt when $k_\star$ is $0.002\,\mathrm{Mpc}^{-1}$.} %a $\sim 2\sigma$ preference for $\betas > 0$ persists. 
We therefore conclude that the preference for blue $\betas$ is stable under the variation of $k_\star$: by studying the improvement in $\chi^2$ %when $$ is left free to vary 
with respect to the $\Lambda\mathrm{CDM}$ and $\Lambda\mathrm{CDM} + \alphas$ models, we can understand what is its origin.}

%--> fix this, and then CDM in conclusions for both here and letter => no, it's too complicated to be done soon enough: I have thousand of other things to do...

%... regarding the addition of $\sum m_\nu$, we do not expect any significant change, since $\sum m_\nu$ and $\betas$ are uncorrelated (see \fig{nrunrun_v_mnu}).%\footnote{This can be seen from the fact that $\sum m_\nu$ has an effect at high multipoles, while the main effect of $\betas$ is changing\dots %--> this can change, exactly, if we change the pivot...}

\begin{table}[!hbtp]
\begin{center}
\begin{tabular}{lcccc}
\toprule
\horsp
$\mathrm{base}$ \vertsp $k_\star = 0.01\,\mathrm{Mpc}^{-1}$ \vertsp $k_\star = 0.002\,\mathrm{Mpc}^{-1}$ \\
\hline
\morehorsp
$n_\mathrm{s}$ \vertsp ${\siround{0.9758}{4}}^{+\siround{0.0117}{4}}_{-\siround{0.0116}{4}}$ \vertsp ${\siround{1.0632}{4}}^{+\siround{0.0466}{4}}_{-\siround{0.0459}{4}}$ \\
\morehorsp
$\alpha_\mathrm{s}$ \vertsp ${\siround{-0.032}{3}}\pm{\siround{0.015}{3}}$ \vertsp ${\siround{-0.076}{3}}\pm{\siround{0.035}{3}}$ \\
\morehorsp
$\beta_\mathrm{s}$ \vertsp ${\siround{0.027}{3}}\pm{\siround{0.013}{3}}$ \vertsp ${\siround{0.027}{3}}\pm{\siround{0.013}{3}}$ \\
\hline
\bottomrule
%\botrule
\end{tabular}
\caption{\footnotesize{$\limit{68}$ constraints on $n_\mathrm{s}$, $\alpha_\mathrm{s}$, $\beta_\mathrm{s}$, for the listed pivot scales: the model is $\Lambda\mathrm{CDM} + \alpha_\mathrm{s} + \betas$, and the dataset is \emph{Planck} ($TT$, $TE$, $EE$ + lowP).}}
\label{tab:pivots}
\end{center}
\end{table}

%\begin{table}[!hbtp]
%\begin{center}
%\begin{tabular}{lcc}
%\toprule
%\horsp
%$\mathrm{base}$ \vertsp $k_\star = 0.01\,\mathrm{Mpc}^{-1}$ \vertsp $k_\star = 0.002\,\mathrm{Mpc}^{-1}$ \\
%\hline
%\morehorsp
%$n_\mathrm{s}$ \vertsp ${\siround{0.9758}{4}}^{+\siround{0.0117}{4}}_{-\siround{0.0116}{4}}$ \vertsp $wait$ \\
%%\morehorsp
%\horsp
%$\alpha_\mathrm{s}$ \vertsp ${\siround{-0.032}{3}}\pm{\siround{0.015}{3}}$ \vertsp $wait$ \\
%%\morehorsp
%\horsp
%$\beta_\mathrm{s}$ \vertsp ${\siround{0.027}{3}}\pm{\siround{0.013}{3}}$ \vertsp $wait$ \\
%%\hline
%%\bottomrule
%\botrule
%\end{tabular}
%\caption{\footnotesize{$\limit{68}$ constraints on $n_\mathrm{s}$, $\alpha_\mathrm{s}$, $\beta_\mathrm{s}$, for the listed pivot scales: the model is $\Lambda\mathrm{CDM} + \alpha_\mathrm{s} + \betas$, and the dataset is \emph{Planck} ($TT$, $TE$, $EE$ + lowP).}}
%\label{tab:pivots}
%\end{center}
%\end{table}

\begin{figure*}[!hbt]
\includegraphics[width=0.48\textwidth]{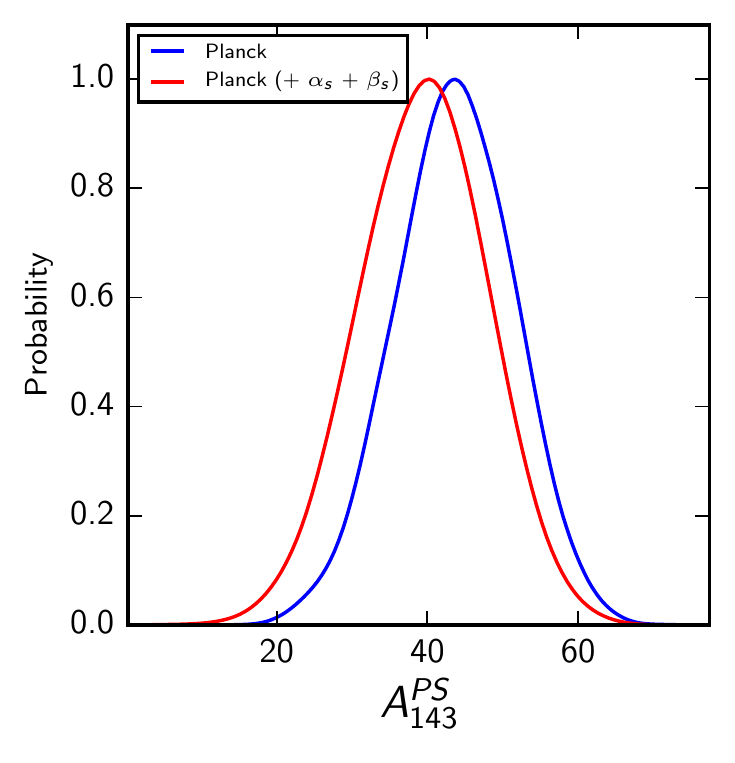}
\caption{\footnotesize{Shift in the amplitude of unresolved foreground point sources
at $143$ GHz between the $\Lambda$CDM case and the case when variation in
$\alphas$ and $\betas$ are considered. The dataset used is \emph{Planck} temperature and
polarization angular spectra.}}
\label{fig:aps143}
\end{figure*}

The \emph{Planck} likelihood consists essentially of three terms: a low-$\ell$ ($\ell=2\div29$) TEB likelihood 
based on the \emph{Planck} LFI $70$ GHz channel full mission dataset, an high-$\ell$ likelihood based on \emph{Planck} HFI
$100$ GHz, $143$ GHz and $217$ GHz channels half mission dataset and, finally, an additional
$\chi^2$ term that comes from the external priors assumed on foregrounds (see \cite{Aghanim:2015xee}).
By looking at the mean $\chi^2_\mathrm{eff}$ values from these three terms we can better understand from where
(low $\ell$, high $\ell$, foregrounds) the indication for $\beta_s$ is coming.
Comparing with the $\chi^2$ values obtained under standard $\Lambda$CDM with
$\alphas=0$ and $\betas=0$, we have found that while the high-$\ell$ likelihood remains
unchanged, there is an improvement in the low-$\ell$ likelihood of $\Delta \chi^2_\mathrm{eff} \sim 2.5$
and in the foregrounds term with $\Delta \chi^2_\mathrm{eff} \sim 1$. The inclusion of $\betas$ provides
therefore a better fit to the low-$\ell$ part of the CMB spectrum and to the foregrounds
prior. While the better fit to the low-$\ell$ part of the CMB spectrum can be easily explained
by the low quadrupole $TT$ anomaly and by the dip at $\ell \sim 20-30$, the change due to foregrounds
is somewhat unexpected since, in general, foregrounds do not correlate with cosmological
parameters. We have found a significant correlation between $\betas$ and the point source amplitude 
at $143$ GHz, $A^{PS}_{143}$. The posterior of $A^{PS}_{143}$ shifts indeed by half sigma towards lower 
values with respect to the standard 
$\Lambda$CDM case (see \fig{aps143}) from $A^{PS}_{143}=43\pm 8$ to 
$A^{PS}_{143}=39\pm8$ at $\limit{68}$. This shift could also explain the small difference between the 
constraints reported here and those reported in \cite{Ade:2015lrj}, that uses 
the \emph{Pliklite} likelihood code where foregrounds are marginalized at their $\Lambda$CDM values.

\textcolor{black}{Before proceeding, we stress that using a likelihood ratio test \cite{Liddle:2004nh} it is easy to see that, for a $\Delta\chi^2_\mathrm{eff}\sim 3.5$ (as the one we find here), there still is a $\sim 17\%$ probability that the $\Lambda\mathrm{CDM}$ model is the correct one.\footnote{\textcolor{black}{Using the fact that $2\log(\mathcal{L}_1/\mathcal{L}_2)$ is distributed as a $\chi^2$ with $\mathrm{d.o.f.} = \mathrm{d.o.f.}_1 - \mathrm{d.o.f.}_2$.}} Given the \emph{Planck} $TT$, $TE$, $EE$ + lowP dataset, this is the significance with which the $\Lambda\mathrm{CDM} + \alphas + \betas$ model is preferred over the $\Lambda\mathrm{CDM}$ one.}

%--> Gigi D'Ag would not be happy...

%--> We stress, however, that our goal was mainly to show how including the running of the running gives a better fit to the low-ell TT power spectrum, even if with a not very strong significance.

%--> http://tex.stackexchange.com/questions/27097/changing-the-font-size-in-a-table

Going back to \tab{results}, we can see that the indication for $\betas>0$ is slightly weakened but still 
present also when external datasets are considered.
Adding CMB lensing gives $\betas=0.022\pm0.013$, \ie reducing the tension to about $1.7$ standard
deviations, while the inclusion of %the WL 
{weak lensing} and BAO {data} does not %significantly degrade 
lead to an appreciable decrease in the statistical significance of 
$\alphas$ and $\betas$.

In Tab.~\ref{tab:results-mnu} we report similar constraints but including also variations in the
neutrino mass absolute scale $\sum m_{\nu}$. The constraints obtained from the \emph{Planck} 2015
data release on the neutrino masses are indeed very strong, especially when combined with 
BAO data, ruling out the possibility of a direct detection from current and future
beta and double beta decay experiments (see, \emph{e.g.}, \cite{gerbino}). Since %the 
\emph{Planck} data show a preference for $\betas>0$, it is clearly interesting to investigate if the inclusion of running has 
some impact on the cosmological constraints on $\sum m_{\nu}$.
Comparing the results of Tab.~\ref{tab:results-mnu} with those in \cite{Ade:2015lrj}, which were obtained
assuming $\alphas=\betas=0$, we see that the constraints on $\sum m_{\nu}$ are only
slightly weakened, moving from $\sum m_{\nu}<0.490\,\mathrm{eV}$ to $\sum m_{\nu}<0.530\,\mathrm{eV}$
at $\limit{95}$ for the \emph{Planck} dataset alone and from $\sum m_{\nu}<0.590\,\mathrm{eV}$ to
$\sum m_{\nu}<0.644\,\mathrm{eV}$ at $\limit{95}$ when also lensing is considered. 
The constraints on $\sum m_{\nu}$ including the WL and BAO datasets are essentially unaffected by $\betas$. We can therefore conclude that there is no
significant correlation between $\betas$ and $\sum m_{\nu}$.

\begin{figure}[!hbt]
%\centering
\includegraphics[width=0.48\textwidth]{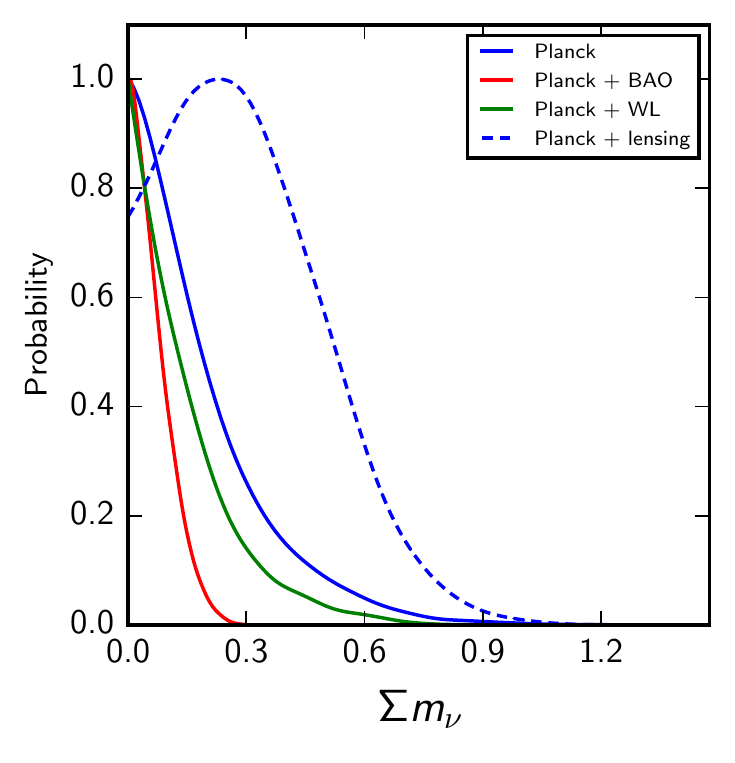}
\caption{\footnotesize{One-dimensional posterior distributions for the sum of neutrino masses $\sum m_\nu$, for the indicated datasets. The model considered is $\Lambda\mathrm{CDM} + \alphas + \betas + \sum m_\nu$.}}
\label{fig:mnu}
\end{figure}

\begin{figure}[!hbt]
%\centering
\includegraphics[width=0.48\textwidth]{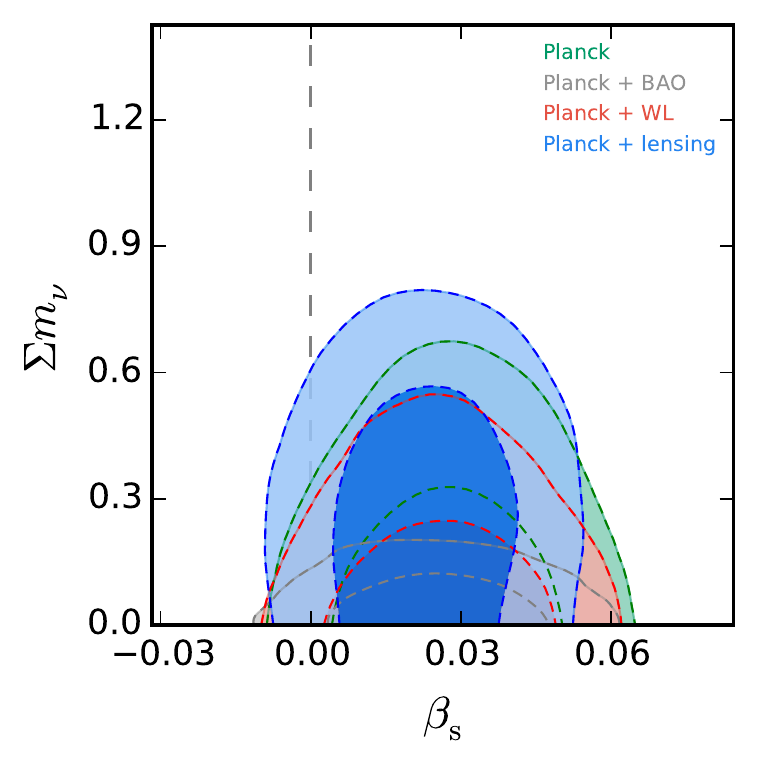}
\caption{\footnotesize{Two-dimensional posteriors in the $\text{$\betas$ -- $\sum m_\nu$}$ plane, for the indicated datasets. We see that there is no correlation between $\sum m_\nu$ and $\betas$.}}
\label{fig:nrunrun_v_mnu}
\end{figure}

In Fig.~\ref{fig:mnu} we plot the posterior distributions for $\sum m_{\nu}$, while in Fig.~\ref{fig:nrunrun_v_mnu} we plot the probability contour at $\limit{68}$ and
$\limit{95}$ for the several combinations of datasets in the $\text{$\betas$ -- $\sum m_{\nu}$}$ plane, respectively.

\begin{table*}[!hbtp]
%\footnotesize
\begin{center}
\begin{tabular}{lcccc}
\toprule
\horsp
$\mathrm{base} + \sum m_\nu$ \vertsp \emph{Planck} \vertsp + lensing \vertsp + WL \vertsp + BAO \\
\hline
\morehorsp
$\Omega_\mathrm{b}h^2$ \vertsp ${\siround{0.02213}{5}}\pm{\siround{0.00018}{5}}$ \vertsp ${\siround{0.02207}{5}}\pm{\siround{0.00019}{5}}$ \vertsp ${\siround{0.02219}{5}}\pm{\siround{0.00018}{5}}$ \vertsp ${\siround{0.02224}{5}}\pm{\siround{0.00015}{5}}$ \\
\morehorsp
$\Omega_\mathrm{c}h^2$ \vertsp ${\siround{0.1208}{4}}\pm{\siround{0.0016}{4}}$ \vertsp ${\siround{0.1206}{4}}\pm{\siround{0.0016}{4}}$ \vertsp ${\siround{0.1199}{4}}\pm{\siround{0.0015}{4}}$ \vertsp ${\siround{0.1196}{4}}\pm{\siround{0.0011}{4}}$ \\
\morehorsp
$100\theta_\mathrm{MC}$ \vertsp ${\siround{1.04062}{5}}^{+\siround{0.00033}{5}}_{-\siround{0.00034}{5}}$ \vertsp ${\siround{1.0406}{5}}\pm{\siround{0.00035}{5}}$ \vertsp ${\siround{1.04072}{5}}\pm{\siround{0.00033}{5}}$ \vertsp ${\siround{1.04082}{5}}\pm{\siround{0.0003}{5}}$ \\
\morehorsp
$\tau$ \vertsp ${\siround{0.095}{3}}^{+\siround{0.019}{3}}_{-\siround{0.02}{3}}$ \vertsp ${\siround{0.08}{3}}\pm{\siround{0.019}{3}}$ \vertsp ${\siround{0.088}{3}}\pm{\siround{0.02}{3}}$ \vertsp ${\siround{0.095}{3}}^{+\siround{0.02}{3}}_{-\siround{0.019}{3}}$ \\
\morehorsp
$\big(\sum m_\nu\big)/\mathrm{eV}$ \vertsp $< \siround{0.53}{3}$ \vertsp $< \siround{0.644}{3}$ \vertsp $< \siround{0.437}{3}$ \vertsp $< \siround{0.159}{3}$ \\
\morehorsp
$H_0$ \vertsp ${\siround{65.76}{2}}^{+\siround{2.12}{2}}_{-\siround{0.99}{2}}$ \vertsp ${\siround{64.76}{2}}^{+\siround{2.49}{2}}_{-\siround{1.7}{2}}$ \vertsp ${\siround{66.46}{2}}^{+\siround{1.76}{2}}_{-\siround{0.91}{2}}$ \vertsp ${\siround{67.38}{2}}\pm{\siround{0.56}{2}}$ \\
\morehorsp
$\log(10^{10} A_\mathrm{s})$ \vertsp ${\siround{3.127}{3}}^{+\siround{0.038}{3}}_{-\siround{0.039}{3}}$ \vertsp ${\siround{3.093}{3}}^{+\siround{0.037}{3}}_{-\siround{0.036}{3}}$ \vertsp ${\siround{3.109}{3}}\pm{\siround{0.038}{3}}$ \vertsp ${\siround{3.124}{3}}^{+\siround{0.037}{3}}_{-\siround{0.038}{3}}$ \\
\morehorsp
$n_\mathrm{s}$ \vertsp ${\siround{0.9576}{4}}^{+\siround{0.0056}{4}}_{-\siround{0.0057}{4}}$ \vertsp ${\siround{0.9583}{4}}\pm{\siround{0.0057}{4}}$ \vertsp ${\siround{0.9601}{4}}^{+\siround{0.0055}{4}}_{-\siround{0.0054}{4}}$ \vertsp ${\siround{0.9612}{4}}^{+\siround{0.0047}{4}}_{-\siround{0.0048}{4}}$ \\
\morehorsp
$\alpha_\mathrm{s}$ \vertsp ${\siround{0.011}{3}}\pm{\siround{0.01}{3}}$ \vertsp ${\siround{0.011}{3}}\pm{\siround{0.01}{3}}$ \vertsp ${\siround{0.012}{3}}\pm{\siround{0.01}{3}}$ \vertsp ${\siround{0.01}{3}}^{+\siround{0.01}{3}}_{-\siround{0.011}{3}}$ \\
\morehorsp
$\beta_\mathrm{s}$ \vertsp ${\siround{0.028}{3}}\pm{\siround{0.013}{3}}$ \vertsp ${\siround{0.023}{3}}\pm{\siround{0.013}{3}}$ \vertsp ${\siround{0.026}{3}}\pm{\siround{0.013}{3}}$ \vertsp ${\siround{0.025}{3}}\pm{\siround{0.013}{3}}$ \\
\hline
\bottomrule
%\botrule
\end{tabular}
\caption{\footnotesize{$\limit{68}$ bounds and $\limit{95}$ upper limits on $\Omega_\mathrm{b}h^2$, $\Omega_\mathrm{c}h^2$, $100\theta_\mathrm{MC}$, $\tau$, $\sum m_\nu$, $H_0$, $\log(10^{10} A_\mathrm{s})$, $n_\mathrm{s}$, $\alpha_\mathrm{s}$, $\beta_\mathrm{s}$, for the listed datasets: the model is $\Lambda\mathrm{CDM} + \alpha_\mathrm{s} + \beta_\mathrm{s} + \sum m_\nu$, $k_\star = 0.05\,\mathrm{Mpc}^{-1}$.}}
\label{tab:results-mnu}
\end{center}
\end{table*}

\begin{table*}[!hbtp]
%\footnotesize
\begin{center}
\begin{tabular}{lcccc}
\toprule
\horsp
$\mathrm{base} + A_L$ \vertsp \emph{Planck} \vertsp + lensing \vertsp + WL \vertsp + BAO \\
\hline
\morehorsp
$\Omega_\mathrm{b}h^2$ \vertsp ${\siround{0.02227}{5}}\pm{\siround{0.00019}{5}}$ \vertsp ${\siround{0.02214}{5}}\pm{\siround{0.00018}{5}}$ \vertsp ${\siround{0.02235}{5}}\pm{\siround{0.00019}{5}}$ \vertsp ${\siround{0.02232}{5}}\pm{\siround{0.00016}{5}}$ \\
\morehorsp
$\Omega_\mathrm{c}h^2$ \vertsp ${\siround{0.1196}{4}}\pm{\siround{0.0017}{4}}$ \vertsp ${\siround{0.1202}{4}}\pm{\siround{0.0017}{4}}$ \vertsp ${\siround{0.1185}{4}}\pm{\siround{0.0016}{4}}$ \vertsp ${\siround{0.119}{4}}\pm{\siround{0.0011}{4}}$ \\
\morehorsp
$100\theta_\mathrm{MC}$ \vertsp ${\siround{1.04081}{5}}\pm{\siround{0.00033}{5}}$ \vertsp ${\siround{1.04076}{5}}\pm{\siround{0.00033}{5}}$ \vertsp ${\siround{1.04093}{5}}\pm{\siround{0.00033}{5}}$ \vertsp ${\siround{1.04089}{5}}\pm{\siround{0.0003}{5}}$ \\
\morehorsp
$\tau$ \vertsp ${\siround{0.07}{3}}\pm{\siround{0.025}{3}}$ \vertsp ${\siround{0.07}{3}}\pm{\siround{0.025}{3}}$ \vertsp $< \siround{0.095}{3}$ \vertsp ${\siround{0.07}{3}}^{+\siround{0.024}{3}}_{-\siround{0.026}{3}}$ \\
\morehorsp
$A_L$ \vertsp ${\siround{1.106}{3}}^{+\siround{0.079}{3}}_{-\siround{0.09}{3}}$ \vertsp ${\siround{0.984}{3}}^{+\siround{0.058}{3}}_{-\siround{0.064}{3}}$ \vertsp ${\siround{1.157}{3}}^{+\siround{0.077}{3}}_{-\siround{0.086}{3}}$ \vertsp ${\siround{1.118}{3}}^{+\siround{0.075}{3}}_{-\siround{0.084}{3}}$ \\
\morehorsp
$H_0$ \vertsp ${\siround{67.38}{2}}\pm{\siround{0.77}{2}}$ \vertsp ${\siround{67.04}{2}}^{+\siround{0.75}{2}}_{-\siround{0.76}{2}}$ \vertsp ${\siround{67.88}{2}}\pm{\siround{0.73}{2}}$ \vertsp ${\siround{67.64}{2}}^{+\siround{0.52}{2}}_{-\siround{0.53}{2}}$ \\
\morehorsp
$\log(10^{10} A_\mathrm{s})$ \vertsp ${\siround{3.073}{3}}^{+\siround{0.05}{3}}_{-\siround{0.051}{3}}$ \vertsp ${\siround{3.074}{3}}^{+\siround{0.05}{3}}_{-\siround{0.051}{3}}$ \vertsp ${\siround{3.044}{3}}^{+\siround{0.044}{3}}_{-\siround{0.051}{3}}$ \vertsp ${\siround{3.072}{3}}\pm{\siround{0.049}{3}}$ \\
\morehorsp
$n_\mathrm{s}$ \vertsp ${\siround{0.9621}{4}}\pm{\siround{0.0062}{4}}$ \vertsp ${\siround{0.9597}{4}}\pm{\siround{0.0061}{4}}$ \vertsp ${\siround{0.9652}{4}}^{+\siround{0.0059}{4}}_{-\siround{0.006}{4}}$ \vertsp ${\siround{0.9637}{4}}\pm{\siround{0.0049}{4}}$ \\
\morehorsp
$\alpha_\mathrm{s}$ \vertsp ${\siround{0.01}{3}}\pm{\siround{0.01}{3}}$ \vertsp ${\siround{0.012}{3}}\pm{\siround{0.01}{3}}$ \vertsp ${\siround{0.01}{3}}\pm{\siround{0.01}{3}}$ \vertsp ${\siround{0.009}{3}}\pm{\siround{0.01}{3}}$ \\
\morehorsp
$\beta_\mathrm{s}$ \vertsp ${\siround{0.021}{3}}\pm{\siround{0.014}{3}}$ \vertsp ${\siround{0.024}{3}}\pm{\siround{0.014}{3}}$ \vertsp ${\siround{0.018}{3}}\pm{\siround{0.013}{3}}$ \vertsp ${\siround{0.019}{3}}\pm{\siround{0.013}{3}}$ \\
\hline
\bottomrule
%\botrule
\end{tabular}
\caption{\footnotesize{$\limit{68}$ bounds and $\limit{95}$ upper limits on $\Omega_\mathrm{b}h^2$, $\Omega_\mathrm{c}h^2$, $100\theta_\mathrm{MC}$, $\tau$, $A_L$, $H_0$, $\log(10^{10} A_\mathrm{s})$, $n_\mathrm{s}$, $\alpha_\mathrm{s}$, $\beta_\mathrm{s}$, for the listed datasets: the model is $\Lambda\mathrm{CDM} + \alpha_\mathrm{s} + \beta_\mathrm{s} + A_L$, $k_\star = 0.05\,\mathrm{Mpc}^{-1}$.}}
\label{tab:results-alens}
\end{center}
\end{table*}

\begin{table*}[!hbtp]
\begin{center}
\begin{tabular}{lcccc}
\toprule
\horsp
$\mathrm{base} + \Omega_K$ \vertsp \emph{Planck} \vertsp + lensing \vertsp + WL \vertsp + BAO \\
\hline
\morehorsp
$\Omega_\mathrm{b}h^2$ \vertsp ${\siround{0.0223}{5}}\pm{\siround{0.00019}{5}}$ \vertsp ${\siround{0.02213}{5}}\pm{\siround{0.00018}{5}}$ \vertsp ${\siround{0.02214}{5}}\pm{\siround{0.00019}{5}}$ \vertsp ${\siround{0.02218}{5}}\pm{\siround{0.00018}{5}}$ \\
\morehorsp
$\Omega_\mathrm{c}h^2$ \vertsp ${\siround{0.1192}{4}}^{+\siround{0.0017}{4}}_{-\siround{0.0018}{4}}$ \vertsp ${\siround{0.1204}{4}}\pm{\siround{0.0017}{4}}$ \vertsp ${\siround{0.1206}{4}}\pm{\siround{0.0017}{4}}$ \vertsp ${\siround{0.1205}{4}}\pm{\siround{0.0016}{4}}$ \\
\morehorsp
$100\theta_\mathrm{MC}$ \vertsp ${\siround{1.04086}{5}}\pm{\siround{0.00034}{5}}$ \vertsp ${\siround{1.04074}{5}}\pm{\siround{0.00033}{5}}$ \vertsp ${\siround{1.04068}{5}}^{+\siround{0.00035}{5}}_{-\siround{0.00034}{5}}$ \vertsp ${\siround{1.04072}{5}}^{+\siround{0.00033}{5}}_{-\siround{0.00034}{5}}$ \\
\morehorsp
$\tau$ \vertsp ${\siround{0.062}{3}}^{+\siround{0.024}{3}}_{-\siround{0.028}{3}}$ \vertsp ${\siround{0.076}{3}}\pm{\siround{0.026}{3}}$ \vertsp ${\siround{0.099}{3}}^{+\siround{0.023}{3}}_{-\siround{0.024}{3}}$ \vertsp ${\siround{0.094}{3}}\pm{\siround{0.018}{3}}$ \\
\morehorsp
$\Omega_K$ \vertsp ${\siround{-0.0302}{4}}^{+\siround{0.025}{4}}_{-\siround{0.0173}{4}}$ \vertsp ${\siround{0.0045}{4}}^{+\siround{0.0096}{4}}_{-\siround{0.0076}{4}}$ \vertsp ${\siround{0.0082}{4}}^{+\siround{0.0091}{4}}_{-\siround{0.0071}{4}}$ \vertsp ${\siround{0.0015}{4}}\pm{\siround{0.0021}{4}}$ \\
\morehorsp
$H_0$ \vertsp ${\siround{57.75}{2}}^{+\siround{4.81}{2}}_{-\siround{6.34}{2}}$ \vertsp ${\siround{69.71}{2}}^{+\siround{4.11}{2}}_{-\siround{4.62}{2}}$ \vertsp ${\siround{71.7}{2}}^{+\siround{3.91}{2}}_{-\siround{5.02}{2}}$ \vertsp ${\siround{67.72}{2}}^{+\siround{0.71}{2}}_{-\siround{0.72}{2}}$ \\
\morehorsp
$\sigma_8$ \vertsp ${\siround{0.799}{3}}^{+\siround{0.033}{3}}_{-\siround{0.036}{3}}$ \vertsp ${\siround{0.837}{3}}\pm{\siround{0.029}{3}}$ \vertsp ${\siround{0.86}{3}}^{+\siround{0.026}{3}}_{-\siround{0.027}{3}}$ \vertsp ${\siround{0.85}{3}}\pm{\siround{0.016}{3}}$ \\
\morehorsp
$\log(10^{10} A_\mathrm{s})$ \vertsp ${\siround{3.057}{3}}^{+\siround{0.048}{3}}_{-\siround{0.058}{3}}$ \vertsp ${\siround{3.087}{3}}\pm{\siround{0.052}{3}}$ \vertsp ${\siround{3.133}{3}}^{+\siround{0.047}{3}}_{-\siround{0.049}{3}}$ \vertsp ${\siround{3.124}{3}}^{+\siround{0.036}{3}}_{-\siround{0.037}{3}}$ \\
\morehorsp
$n_\mathrm{s}$ \vertsp ${\siround{0.9642}{4}}^{+\siround{0.0064}{4}}_{-\siround{0.0065}{4}}$ \vertsp ${\siround{0.9589}{4}}^{+\siround{0.0064}{4}}_{-\siround{0.0063}{4}}$ \vertsp ${\siround{0.9574}{4}}\pm{\siround{0.0063}{4}}$ \vertsp ${\siround{0.9587}{4}}\pm{\siround{0.0057}{4}}$ \\
\morehorsp
$\alpha_\mathrm{s}$ \vertsp ${\siround{0.008}{3}}^{+\siround{0.01}{3}}_{-\siround{0.011}{3}}$ \vertsp ${\siround{0.013}{3}}\pm{\siround{0.01}{3}}$ \vertsp ${\siround{0.014}{3}}\pm{\siround{0.011}{3}}$ \vertsp ${\siround{0.011}{3}}\pm{\siround{0.01}{3}}$ \\
\morehorsp
$\beta_\mathrm{s}$ \vertsp ${\siround{0.013}{3}}\pm{\siround{0.014}{3}}$ \vertsp ${\siround{0.027}{3}}^{+\siround{0.015}{3}}_{-\siround{0.017}{3}}$ \vertsp ${\siround{0.035}{3}}^{+\siround{0.015}{3}}_{-\siround{0.017}{3}}$ \vertsp ${\siround{0.027}{3}}\pm{\siround{0.014}{3}}$ \\
\hline
\bottomrule
%\botrule
\end{tabular}
\caption{\footnotesize{$\limit{68}$ bounds on $\Omega_\mathrm{b}h^2$, $\Omega_\mathrm{c}h^2$, $100\theta_\mathrm{MC}$, $\tau$, $\Omega_K$, $H_0$, $\sigma_8$, $\log(10^{10} A_\mathrm{s})$, $n_\mathrm{s}$, $\alpha_\mathrm{s}$, $\beta_\mathrm{s}$, for the listed datasets: the model is $\Lambda\mathrm{CDM} + \alpha_\mathrm{s} + \beta_\mathrm{s} + \Omega_K$, $k_\star = 0.05\,\mathrm{Mpc}^{-1}$.}}
\label{tab:results-omegak}
\end{center}
\end{table*}

In Tab. \ref{tab:results-alens} we report the constraints from the same datasets but letting 
also the lensing amplitude $A_{L}$ free to vary. As discussed in the introduction, 
\emph{Planck} data are also suggesting a value for $A_{L}>1$ and is therefore 
interesting to check if there is a correlation with $\betas$. 
As we can see there is a correlation between the two parameters but not extremely significant. 
Even with a lower statistical significance, at about $\sim\text{$1.2$ -- $1.5$}$ standard deviations for $A_{L}$ and $\beta_\mathrm{s}$ respectively (that could be also explained by the increased volume of parameter space), data seem
to suggest the presence of {\it both} anomalies.
When the CMB lensing data are included, $A_{L}$ goes back to its standard value while
the indication for $\betas$ increases. When the WL shear data are included the 
$A_{L}$ anomaly is present while the indication for $\betas$ is weakened.

We also consider variation in the curvature of the universe and we report the constraints
in Tab.~\ref{tab:results-omegak}. 
As we can see, also in this case we have a correlation between $\betas$ and
$\Omega_k$ but not significant enough to completely cancel any indication
for these anomalies from \emph{Planck} data. Indeed, when $\Omega_k$ is considered,
we have still a preference for $\Omega_k<0$ and $\betas>0$ at more than one standard
deviation. More interestingly, when external datasets are included, the indication
for a positive curvature simply vanishes, while we get $\betas>0$ slightly below $\limit{95}$.

\begin{figure*}
\begin{center}
\begin{tabular}{c c}
\includegraphics[width=\columnwidth]{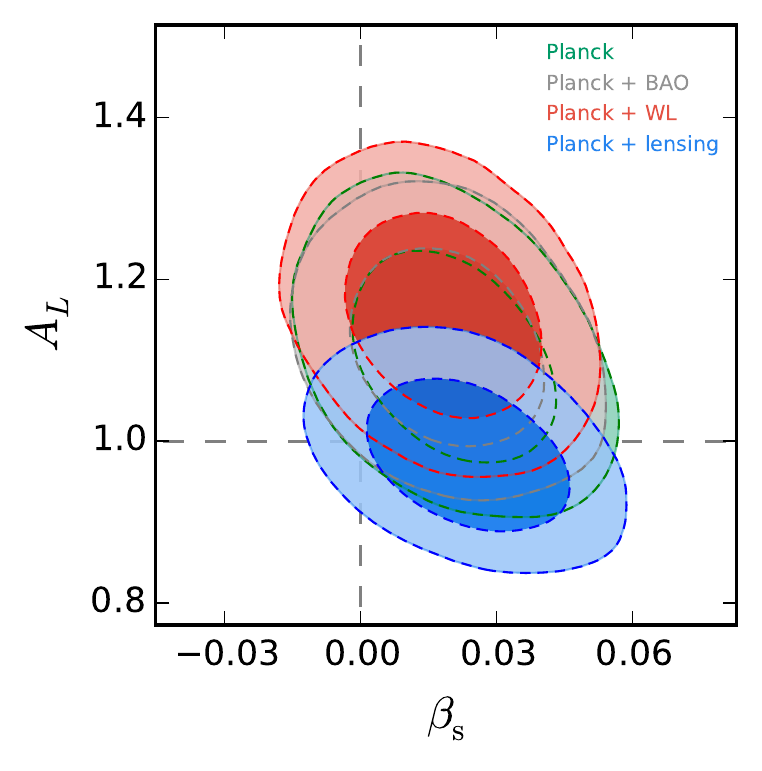}
&\includegraphics[width=\columnwidth]{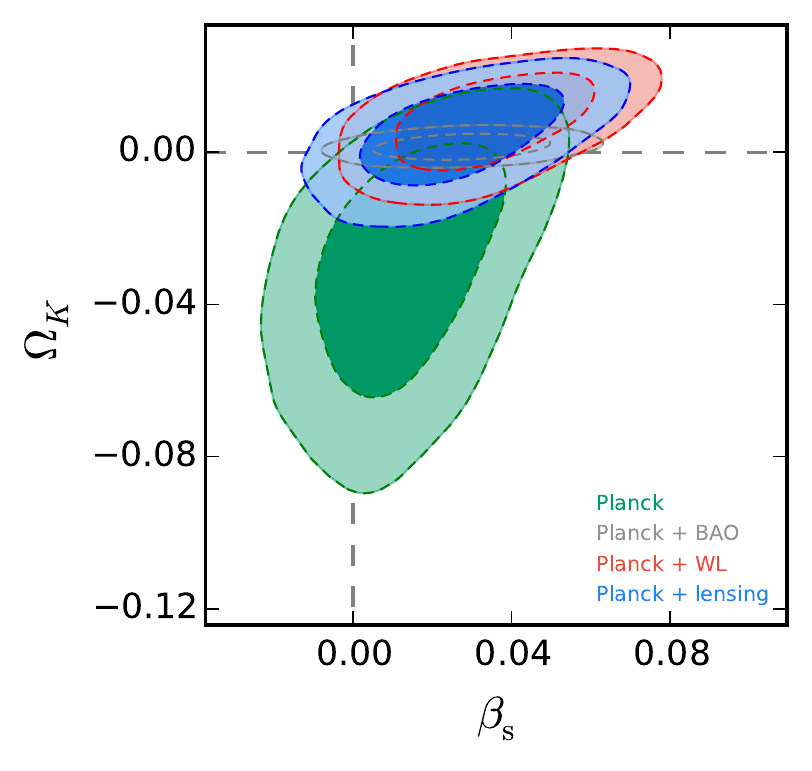}
\end{tabular}
\end{center}
\caption{\footnotesize{Constraints at $\limit{68}$ and $\limit{95}$ in the $\text{$\betas$ -- $A_{L}$}$ plane (left panel)
and in the $\text{$\betas$ -- $\Omega_k$}$ plane (right panel).}}
\label{fig:alens_omegak-v-nrunrun}
\end{figure*}

In Fig. \ref{fig:alens_omegak-v-nrunrun} we show the constraints at $\limit{68}$ and $\limit{95}$ in the $\text{$\betas$ -- $A_{L}$}$ plane (left panel)
and in the $\text{$\betas$ -- $\Omega_k$}$ plane (right panel).

%%%% Ly-\alpha %%%%

%\tred{\textbf{On Ly-$\alpha$}: while constraints on the scale dependence of the primordial power spectrum from observations of the Ly-$\alpha$ forest could be very powerful (since the forest constrains wavenumbers $k\approx 1h\,\mathrm{Mpc}^{-1}$, greatly extending the lever arm w.r.t. CMB observations), modeling the ionization state and thermodynamic properties of the intergalactic medium to convert flux measurements into a density power spectrum is very challenging (see \cite{Zaldarriaga:2000mz} and \cite{Adshead:2010mc} for a discussion). For this reason we do not investigate this topic further in this work.}\footnote{\tred{We refer to \cite{Palanque-Delabrouille:2015pga} for a recent analysis of one-dimensional Ly-$\alpha$ forest power spectrum measured in \cite{Palanque-Delabrouille:2013gaa}, that provides also small-scale constraints on the tilt $\ns$ and the running $\alphas$.}} 

%%%% Ly-\alpha %%%%

%\section{Implications for Galaxy Formation and Primordial Black Holes}
%\%label{sec:galaxies+BHs}

\textcolor{black}{We conclude this section by looking at what are the improvements (or non-improvements) in $\chi^2$ over our base model $\Lambda\mathrm{CDM} + \alphas + \betas$ when additional parameters ($A_L$, $\sum m_\nu$ and $\Omega_K$) are considered: the tables (Tabs.~\ref{tab:chi2-pl}, \ref{tab:chi2-pl+lens}, \ref{tab:chi2-pl+wl} and \ref{tab:chi2-pl+bao}) containing all the $\Delta\chi^2$ (which we define by $\chi^2_\mathrm{base} - \chi^2_\text{base + ext.}$) are collected in \sect{chi2}. When considering the ${} + A_L$ extension, we see that an improvement $\Delta\chi^2\sim 1.5$ ($\Delta\chi^2\sim 6$) is obtained for the \emph{Planck} $TT$, $TE$, $EE$ + lowP + BAO dataset (\emph{Planck} $TT$, $TE$, $EE$ + lowP + WL dataset), while the addition of CMB lensing data to \emph{Planck} temperature and polarization data leads to $\Delta\chi^2\sim-1.5$, mainly driven by a worse fit to the foregrounds.} %--> basically Alens goes back towards one when we include lensing data.
\textcolor{black}{When $\sum m_\nu$ or $\Omega_K$ are left free to vary, we see that the fit to the data is in general worse: only when adding $\Omega_K$ to the \emph{Planck} $TT$, $TE$, $EE$ + lowP + WL dataset we get a $\Delta\chi^2\sim 2$ improvement.}

%--> add something like "mainly driven by the abc part of the likelihood". Add line with the whole \Delta\chi^2: call it \Delta\chi^2, remove first line in left...

\section{Present and Future Constraints from $\mu$-Distortions}
\label{sec:distortions}

\noindent \textcolor{black}{CMB $\mu$-type spectral distortions \cite{Zeldovich:1969ff, Sunyaev:1970er} from the dissipation of acoustic waves at redshifts between $z = \num{2d6}\equiv z_\mathrm{dC}$ and $z = \num{5d4}\equiv z_{\mu\text{-}y}$ offer a window on the primordial power spectrum at very small scales, ranging from $50$ to $\num{d4}$ $\mathrm{Mpc}^{-1}$ (for most recent works on this topic see \cite{Dent:2012ne, Chluba:2012we, Khatri:2013dha, Clesse:2014pna, Enqvist:2015njy, Cabass:2016giw, Chluba:2016bvg} and references therein). The impact of a PIXIE-like mission on the constraints on the running $\alphas$ has been recently analyzed in \cite{Cabass:2016giw}, while \cite{Khatri:2013dha, Chluba:2016bvg} also investigated the variety of signals (and corresponding forecasts) that are expected in the $\Lambda$CDM model (not limited to a $\mu$-type distortion).}

In this section, we briefly investigate the constraining power of $\mu$-distortions on $\betas$, given the \emph{Planck} constraints on $\alphas$ and $\betas$ of \sect{results}. We %use 
compute the contribution to the $\mu$-monopole from Silk damping of acoustic waves in the photon-baryon plasma \cite{Silk:1967kq, Peebles:1970ag, Kaiser:1983abc, Hu:1992dc, Chluba:2012gq}, using the expression for the distortion visibility function presented in \cite{Khatri:2013dha}.\footnote{This is related to the method called ``Method II'' in \cite{Chluba:2016bvg}, the difference being the visibility function $J_\mathrm{bb}(z)$ used: $J_\mathrm{bb}(z)$ is approximated to $\exp(-(z/z_\mathrm{dC})^{5/2})$ in the ``Method II'' of \cite{Chluba:2016bvg}, while \cite{Khatri:2013dha} derives a fitting formula to take into account the dependence of $J_\mathrm{bb}(z)$ on cosmological parameters. At the large values of $\alphas$ and $\betas$ allowed by \emph{Planck}, we do not expect this difference to be very relevant for our final result.} To understand the relationship between the $\mu$ amplitude and the parameters of the primordial power spectrum, one can compute the (integrated) fractional energy that is dissipated by the acoustic waves $\delta_\gamma$ between $z = \num{2d6}$ and $z = \num{5d4}$: this energy feeds back into the background and generates $\mu$-distortions according to (see also \cite{Pajer:2012vz, Pajer:2013oca}) %--> J_bb depends on Y_P and omegabh2, for starters...
\begin{equation}
\label{eq:mu-silk}
%\begin{split}
\mu(\vec{x})
%&
\approx\frac{1.4}{4}\braket{\delta_\gamma^2(z,\vec{x})}_p\Big|^{z_\mathrm{dC}}_{z_{\mu\text{-}y}} %\\
%&\approx 2.3\int\frac{\dif\vec{k}_1 \dif\vec{k}_2}{(2\pi)^3}e^{i\vec{k}_+\cdot\vec{x}}\zeta_{\vec{k}_1}\zeta_{\vec{k}_2} 
%e^{-(k^2_1 + k^2_2)/k^2_D}\Big|^{z_\mathrm{dC}}_{z_{\mu\text{-}y}}
\,\,, 
%\end{split}
\end{equation}
where $\braket{\dots}_p$ indicates the average over a period of oscillation and $\zeta$ is the primordial curvature perturbation. The diffusion damping length appearing in the above formula is given by \cite{Silk:1967kq, Peebles:1970ag, Kaiser:1983abc}
\begin{equation}
\label{eq:diff-dist}
k_\tu{D}(z) = \sqrt{\int^{+\infty}_z\dif z\frac{1+z}{H n_e\sigma_\tu{T}}\bigg[\frac{R^2 + \frac{16}{15}(1+R)}{6(1+R)^2}\bigg]}\,\,.
\end{equation}
The observed $\mu$-distortion monopole is basically the ensemble average of $\mu(\vec{x})$ at $z = \num{5d4}$: by averaging \eq{mu-silk}, then, one sees that it is equal to the log-integral of the primordial power spectrum multiplied by a window function 
\begin{equation}
\label{eq:mu-window}
W_\mu(k) = 2.3\,e^{-2k^2/k^2_D}\Big|^{z_\mathrm{dC}}_{z_{\mu\text{-}y}}\,\,,
\end{equation}
which localizes the integral between $50\,\mathrm{Mpc}^{-1}$ to $\num{d4}\,\mathrm{Mpc}^{-1}$.

\begin{figure*}
\begin{center}
\begin{tabular}{c c}
\includegraphics[width=\columnwidth]{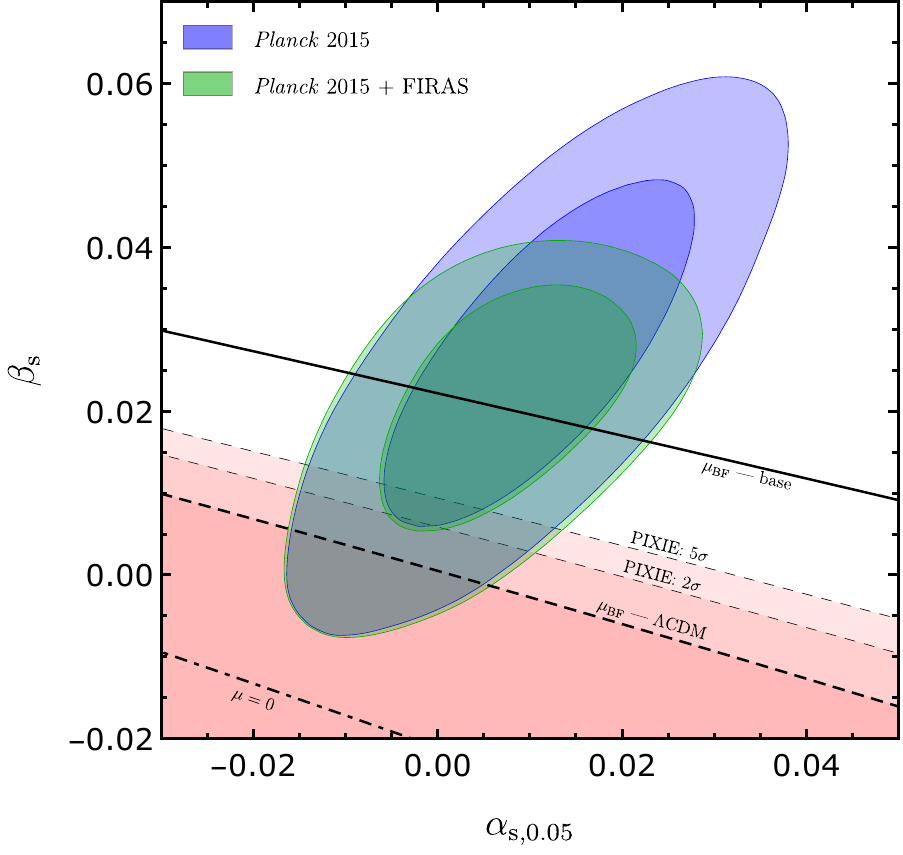}
&\includegraphics[width=\columnwidth]{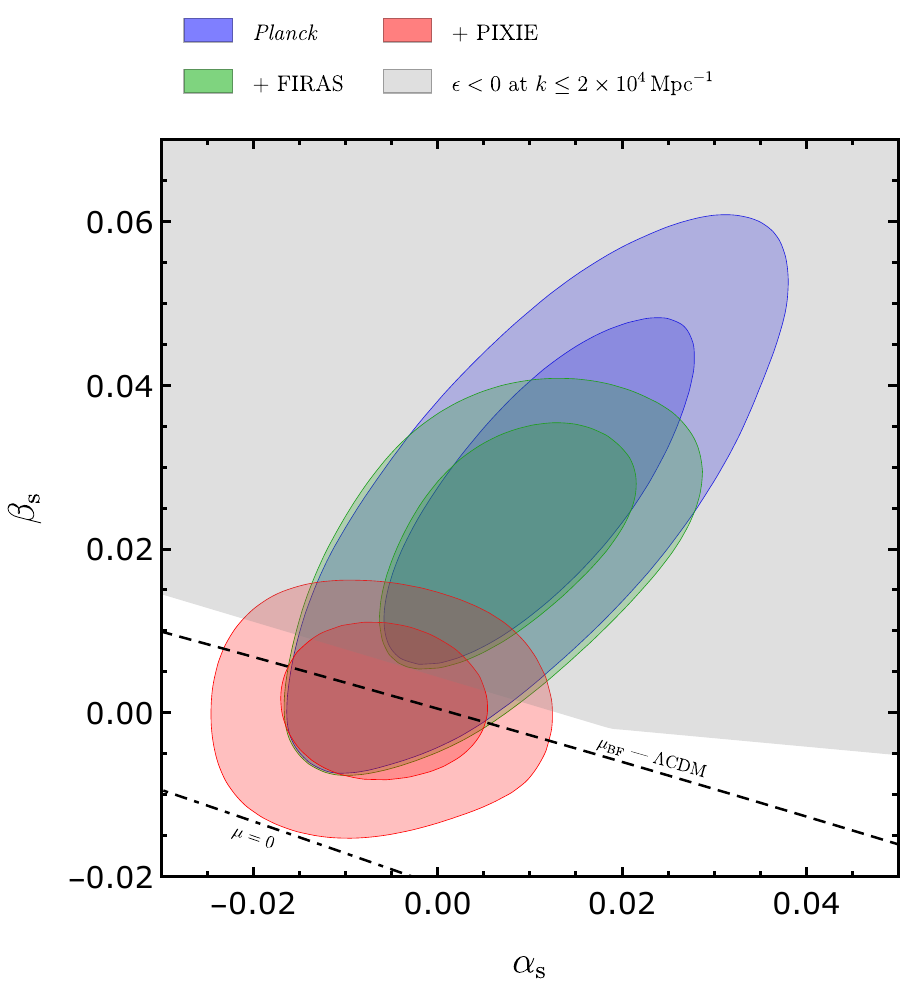}
\end{tabular}
\end{center}
\caption{\footnotesize{Left panel: $\limit{68}$ and $\limit{95}$ contours in the $\alphas$ -- $\betas$ plane, for the \emph{Planck} (blue) and \emph{Planck} + FIRAS (green) datasets (base model). The red regions represent the $2\sigma$ and $5\sigma$ limits from PIXIE around the \emph{Planck} best-fit for the $\Lambda\mathrm{CDM}$ model, \ie $\mu = \num{1.57d-8}$ \cite{Cabass:2016giw}. Right panel: same as left panel, with the red contours represent the $\limit{68}$ and $\limit{95}$ limits from PIXIE, obtained by post-processing the Markov chains with a Gaussian likelihood $\mu = (1.57\pm1.00)\times\num{d-8}$. The grey region represents the values of $\alphas$ and $\betas$ that lead to a slow-roll parameter $\epsilon(k)$, computed via the Taylor expansion of \eq{epsilon_SR}, less than zero before or at $k = \num{2d4}\,\mathrm{Mpc}^{-1}$.}}
\label{fig:theo_bound}
\end{figure*}

\tab{results-mu} shows how, already with the current limit on the $\mu$-distortion amplitude from the FIRAS instrument on the COBE satellite, namely $\mu = (1 \pm 4)\times 10^{-5}$ at $\limit{68}$ \cite{Fixsen:1996nj}, we can get a $28\%$ increase in the $\limit{95}$ upper limits on $\alphas$, and a $33\%$ increase in those on $\betas$ (we also stress that, in the case of $\betas$ fixed to zero, including FIRAS does not result in any improvement on the bounds for $\alphas$). In \fig{theo_bound}, we also report a {forecast for PIXIE}, whose expected error on $\mu$ is $\num{d-8}$ \cite{Kogut:2011xw}.\footnote{In \cite{Chluba:2013pya} it was shown that, when also $r$-distortions are considered, the expected error should be larger (about $\sigma_\mu = \num{1.4d-8}$): however at the large values of $\alphas$ and $\betas$ allowed by \emph{Planck}, %this makes hardly a difference 
the forecasts of \tab{results-mu} are not significantly affected. \textcolor{black}{$r$-distortions are the residual distortions that encode the information on the transition between the $\mu$-era (when distortions are of the $\mu$-type) and the $y$-era (when the CMB is not in kinetic equilibrium and energy injections result in distortions of the $y$-type). We refer to \cite{Khatri:2012tw, Chluba:2013vsa} for a study of these residual distortions, and to \cite{Khatri:2013dha,Chluba:2013pya} for a study of their constraining power on cosmological parameters.}} Besides, we see that:
\begin{itemize}[leftmargin=*]
\item for the best-fit values of cosmological parameters in the $\Lambda\mathrm{CDM} + \alphas + \betas$ model, which leads to $\mu = \num{1.09d-6}$, PIXIE will be able to detect spectral distortions from Silk damping at extremely high significance (\fig{theo_bound}). Besides, we see that {a statistically significant detection of $\betas$ is expected}, %\footnote{{We leave a more quantitative forecasts to future work.}} 
along with a {sizable shrinking} of the {available parameter space} (\fig{theo_bound}). As we discuss later, any detection of such values of $\mu$-distortions will rule out single-field slow-roll inflation, if we assume that all the generated distortions are due to Silk damping and not to other mechanisms like, for example, decaying Dark Matter particles;\footnote{We did not investigate, in this work, whether it could be possible to have models of multi-field inflation (or models where the slow-roll assumption is relaxed \cite{Destri:2008fj}) that can predict such values for the $\mu$-distortion amplitude. \textcolor{black}{We refer to \cite{Clesse:2014pna} for an analysis of some multi-field scenarios.}}
\item for a fiducial value of $\mu$ corresponding to the $\Lambda\mathrm{CDM}$ best-fit %(at the edge of the $2\sigma$ contours of \emph{Planck}), 
\ie $\mu = \num{1.57d-8}$ \cite{Cabass:2016giw}, we see that 
%values of $\betas$ larger than $0.02$ will be excluded at $\sim 5\sigma$. 
we get a $84\%$ increase in the $\limit{95}$ upper limits on $\alphas$, and a $83\%$ increase in those on $\betas$. More precisely, values of $\betas$ larger than $0.02$ will be excluded at $\sim 5\sigma$.
\end{itemize}

\begin{table}[!hbt]
\begin{center}
\begin{tabular}{lccc}
\toprule
\horsp
base \vertsp $\alphas$ \vertsp $\betas$ \vertsp $\mu$ \\
\hline
\morehorsp
\emph{Planck} \vertsp $0.011\pm0.021$ \vertsp $0.027\pm0.027$ \vertsp $/$ \\ %--> 0.0108277 + 0.0205128 - 0.0207688 || 0.026857 + 0.0255371 - 0.02619 || ...
\horsp
%\emph{Planck} 
+ FIRAS \vertsp $0.006^{+0.017}_{-0.018}$ \vertsp $0.020^{+0.016}_{-0.019}$ \vertsp $(0.77^{+3.10}_{-0.77})\times\num{d-6}$ \\ 
%--> 0.00615381 + 0.0167617 - 0.0175234 || 0.019605 + 0.0156703 - 0.0188863 || 7.70191 x 10^-6 + 3.10126 x 10^-5 - 7.70348 x 10^-6
\horsp
%\emph{Planck} 
+ PIXIE \vertsp $-0.007^{+0.012}_{-0.013}$ \vertsp $0.001^{+0.008}_{-0.009}$ \vertsp $(1.59^{+1.75}_{-1.52})\times\num{d-8}$ \\ 
%--> -0.00645864 + 0.0124826 - 0.0132628 || 0.000858908 + 0.00780676 - 0.00886876 || 1.59142 x 10^-8 + 1.51993 x 10^-8 - 1.74866 x 10^-8
\botrule
\end{tabular}
\caption{\footnotesize{$\limit{95}$ bounds on $\alphas$ and $\betas$ from the \emph{Planck} ($TT$, $TE$, $EE$ + lowP), \emph{Planck} + FIRAS and \emph{Planck} + PIXIE datasets, for the $\Lambda\mathrm{CDM} + \alphas + \betas$ (\ie ``base'') model. The results have been obtained by post-processing with a Gaussian likelihood the Markov chains considering $\mu =(1.0\pm4.0)\times10^{-5}$ \cite{Fixsen:1996nj} for FIRAS, and $\mu = (1.57\pm1.00)\times\num{d-8}$ for PIXIE. See the main text for a discussion of the bounds on the $\mu$-amplitude.}}
\label{tab:results-mu}
\end{center}
\end{table}

%\begin{figure}[!hbt]
%%\centering
%\includegraphics[width=.48\textwidth]{nrun_v_nrunrun-work_in_progress-full-no_50_sigma.pdf}
%\caption{\footnotesize{$\limit{68}$ and $\limit{95}$ contours in the $\alphas$ -- $\betas$ plane, for the \emph{Planck} (blue) and \emph{Planck} + FIRAS (green) datasets (base model). %The orange region around the best-fit of the \emph{Planck} analysis ($\mu = \num{1.09d-6}$) represents the $50\sigma$ limits on $\mu$ from PIXIE, taking $\sigma_\mu = \num{1.0d-8}$ from \cite{Kogut:2011xw}. 
%The red regions represent the $2\sigma$ and $5\sigma$ limits from PIXIE around the \emph{Planck} best-fit for the $\Lambda\mathrm{CDM}$, \ie $\mu = \num{1.57d-8}$ \cite{Cabass:2016giw}.}}
%\label{fig:figure-mu}
%\end{figure}
%
%
%\begin{figure}[!hbt]
%%\centering
%\includegraphics[width=.48\textwidth]{nrun_v_nrunrun-work_in_progress-full-no_50_sigma-pixie_lcdm_bf.pdf}
%\caption{\footnotesize{%$\limit{68}$ and $\limit{95}$ contours in the $\alphas$ -- $\betas$ plane, for the \emph{Planck} (blue) and \emph{Planck} + FIRAS (green) datasets (base model).
%Same as \fig{figure-mu}, with the constraints from post-processing Markov chains with a Gaussian likelihood for PIXIE ($\mu = (1.57\pm1.00)\times\num{d-8}$).}}
%\label{fig:figure-mu_with_pixie}
%\end{figure}

We conclude this section with a comment on the validity of a Taylor expansion (in $\log k/k_\star$) of the power spectrum down to scales probed by spectral distortions. We can estimate the terms in the expansion of $n_\mathrm{s}(k)$ by choosing $k= 10^4$ Mpc$ ^{-1} $, corresponding to $k_\mathrm{D}$ at $z= z_\mathrm{dC}$: for values of $\betas$ of order $0.06$ (which are still allowed at $\limit{95}$, as shown in \fig{theo_bound}), the term $\frac{\betas}{6}(\log k/k_\star)^2$ in Eq. (\ref{eq:Delta_of_k}) becomes of order $1$. For this reason, \tab{results-mu} does not report the %current 
limits on $\mu$ coming from the current \emph{Planck} constraints on the scale dependence of the spectrum. When existing limits on $\mu$ from FIRAS are instead added, an extrapolation of $\Delta_\zeta^2(k)$ at the scales probed by $\mu$-distortions starts to become meaningful, and when also PIXIE is included in our forecast around the $\Lambda\mathrm{CDM}$ prediction, the upper bounds on $\alphas$ and $\betas$ are lowered enough that a perturbative expansion becomes viable, %thereby 
making our forecast valid. %--> this is not really the same as using the fact that temperature anisotropies are small to argue that we can use linear theory, because it is a forecast and depends on the fiducial.

%--> In this sense we interpret these bounds from small scale observables as necessary to make sense of the bounds from Planck.

\section{Large $\betas$ and Slow-Roll Inflation}
\label{sec:slow_roll_inflation}

\noindent In this section we discuss briefly the implications that values of $\alphas$ and $\betas$ of order $\num{d-2}$ have for slow-roll inflation. We can compute the running of the slow-roll parameter $\epsilon$ in terms of $\ns$, $\alphas$ and $\betas$ by means of the simple slow-roll relations
\begin{subequations}
\label{eq:slow_roll_rels}
\begin{align}
&N - N_\star = -\log\frac{k}{k_\star}\,\,, \label{eq:slow_roll_rels-1} \\
&1 - \ns = 2\epsilon-\frac{1}{\epsilon}\frac{\dif\epsilon}{\dif N} %- \frac{c_{\mathrm{s},N}}{c_\mathrm{s}}
\,\,, \label{eq:slow_roll_rels-2}
\end{align}
\end{subequations}
where $N$ is the number of e-foldings from the end of inflation, decreasing as time increases (\ie $H\dif t=-\dif N$), and \eq{slow_roll_rels-1} holds we if neglect the time derivative of the inflaton speed of sound $c_\mathrm{s}$. The running of $\epsilon$ up to third order in $N$ is given, then, by %(see \sect{appendix-slow_roll}) %--> this is not an expansion in 1/N. It is just a Taylor series for \epsilon...
\begin{equation}
\label{eq:epsilon_SR}
\epsilon(N) = \epsilon(N_\star) + \sum_{i = 1}^3\frac{\epsilon^{(i)}}{i!}(N-N_\star)^i\,\,, %--> N-N_\star = \pm\log k/k_\star => no, quello e' comunque \epsilon a N_\star...
\end{equation}
where the coefficients $\epsilon^{(i)}$ are given in \sect{appendix-slow_roll}. 

%By plugging in $\ns = 0.96$, $\alphas = 0.01$ and $\betas = 0.02$, taking $\epsilon_\star = 0.002$ (\ie the maximum value allowed by current bounds on $r$, when %the inflaton speed of sound 
%$c_\mathrm{s}$ is fixed to $1$) and using \eq{slow_roll_rels-2}, we see that $\epsilon$ becomes negative at $k\approx 160\,\mathrm{Mpc}^{-1}$.

%This tells us that, even if a 
%single-field slow-roll model can predict $\alphas\approx\betas\approx\mathcal{O}(\num{d-1})$ at CMB scales, it will difficult for it to maintain the flatness of the approximate scale invariance of the spectrum at small scales. 

At scales around $k_\star$, $\ns$ dominates, so that $\epsilon$ is increasing and a red spectrum is obtained. However, in presence of positive $\alphas$ and $\betas$, at small scales $\epsilon$ becomes smaller, until it becomes zero at $k\approx 39.7\,\mathrm{Mpc}^{-1}$ for $\alphas = 0.01$ and $\betas = 0.02$ (taking $\epsilon_\star = 0.002$, \ie the maximum value allowed by current bounds on $r$, when the inflaton speed of sound 
$c_\mathrm{s}$ is fixed to $1$). If we impose that $\epsilon$ stays positive down to $k\approx\num{2d4}\,\mathrm{Mpc}^{-1}$, which is of the same order of magnitude of the maximum $k$ probed by $\mu$-distortions (see \sect{distortions}), %then 
we can obtain a theoretical bound on $\alphas$ and $\betas$. We show this bound in \fig{theo_bound}: this plot tells us that a large part of the contours from \emph{Planck} + FIRAS and \emph{Planck} + PIXIE cannot be interpreted in the context of slow-roll inflation extrapolated to $\mu$-distortion scales, because $\epsilon$ becomes negative before reaching $k\approx k_\mathrm{D}(z_\mathrm{dC})$.\footnote{We point out that it is possible to obtain large positive $\alphas$ and $\betas$ in slow-roll inflation when modulations of the potential are considered \cite{Kobayashi:2010pz}. However, we will not consider such models in this work.}

A similar discussion was presented in \cite{Powell:2012xz}: by means of a slow-roll reconstruction of the inflaton potential \cite{Liddle:1994dx, Kinney:2002qn}, it was shown that if $\betas$ is controlled only by leading-order terms in the slow-roll expansion (see \sect{appendix-slow_roll}), it is not possible to find a single-field inflation model that fits the posteriors from \emph{Planck}.

%\begin{figure}[!hbt]
%%\centering
%\includegraphics[width=.48\textwidth]{nrun_v_nrunrun-work_in_progress-theo_bound-full.pdf}
%\caption{\footnotesize{$\limit{68}$ and $\limit{95}$ contours in the $\alphas$ -- $\betas$ plane, for the \emph{Planck} (blue) and \emph{Planck} + FIRAS (green) datasets (base model). The red regions represent the $\limit{68}$ and $\limit{95}$ limits from PIXIE around the \emph{Planck} best-fit for the $\Lambda\mathrm{CDM}$, \ie $\mu = \num{1.57d-8}$ \cite{Cabass:2016giw}, obtained by post-processing Markov chains with a Gaussian likelihood $\mu = (1.57\pm1.00)\times\num{d-8}$. The grey region represents the values of $\alphas$ and $\betas$ that lead to a $\epsilon(k)$, computed via the Taylor expansion of \eq{epsilon_SR}, less than zero before or at $k = \num{2d4}\,\mathrm{Mpc}^{-1}$.}}
%\label{fig:theo_bound}
%\end{figure}

These kind of bounds tell us that the Taylor expansion is not suitable for extrapolating the inflationary spectrum far away from the CMB window, in presence of the values of $\alphas$ and $\betas$ that are currently allowed by \emph{Planck}, since $\epsilon$ becomes zero already $\sim 7$ e-folds after the horizon exit of $k_\star$. To avoid this problem, one could consider a series expansion that takes into account the theoretical bounds on $\epsilon$, \ie $\epsilon(N = 0) = 1$ and $0 < \epsilon < 1$: %\footnote{With the option of considering $\dif\epsilon/\dif N > 0$.} 
the Taylor series does not respect these requirements, so it does \textit{not} in general represent a possible power spectrum from inflation, over the whole range of scales. Only when the values of the phenomenological parameters describing the scale dependence of the spectrum are small, %the validity 
the Taylor expansion can be a good approximation of a realistic power spectrum over a range of scales much larger than those probed by the CMB.

%--> it is N-Nstar that must be small, i.e. log(k/k_star)... So there is not a "60", but a "60" minus something...

Another possibility is to consider bounds on the primordial power spectrum coming from observables that lie outside the CMB scales, but are still at small enough $k$ that the Taylor series applies. These would be complementary to spectral distortions, which are basically sensitive only to scales around $740\,\mathrm{Mpc}^{-1}$ \cite{Chluba:2012we, Chluba:2013pya}, opening the possibility of multi-wavelength constraints on the scale dependence of the spectrum.

In this regard, %constraints on the scale dependence of the primordial power spectrum from 
observations of the Ly-$\alpha$ forest could be very powerful (%since 
the forest constrains wavenumbers $k\approx 1h\,\mathrm{Mpc}^{-1}$%, greatly extending the lever arm w.r.t. CMB observations
),\footnote{Even if modeling the ionization state and thermodynamic properties of the intergalactic medium to convert flux measurements into a density power spectrum is very challenging (see \cite{Zaldarriaga:2000mz} and \cite{Adshead:2010mc} for a discussion).} In \cite{Palanque-Delabrouille:2015pga}, an analysis of the the one-dimensional Ly-$\alpha$ forest power spectrum measured in \cite{Palanque-Delabrouille:2013gaa} was carried out, showing that it provides also small-scale constraints on the tilt $\ns$ and the running $\alphas$: more precisely, for a $\Lambda\mathrm{CDM} + \alphas + \sum m_{\nu}$ model, a detection at approximately $3\sigma$ of $\alphas$ ($\alphas = -0.00135^{+0.0046}_{-0.0050}$ at $\limit{68}$) is obtained. It would be interesting to carry out this analysis including the running of the running, to see if the bounds on $\betas$ are also lowered. %--> There was also mnu, but there is no correlation between the two? Maybe yes, be careful => remember the smaller scales of free streaming...

\section{Conclusions}
\label{sec:conclusions}

In this paper we have presented new constraints on the running of the running of the scalar spectral index $\betas$ and
discussed in more detail %its anomalous value.
the $2\sigma$ indication for $\betas > 0$ that comes from the analysis of CMB anisotropies data from the \emph{Planck} satellite. 

We have extended previous analyses by considering simultaneous variations in the lensing amplitude parameter
$A_{L}$ and the curvature of the universe $\Omega_k$. 
We have found that, while a correlation does exist between these parameters, \emph{Planck} data still hint for non-standard
values in the extended $\Lambda\mathrm{CDM} +\alphas+\betas+A_L$ and $\Lambda\mathrm{CDM} +\alphas+\betas+\Omega_k$ model, %when the analysis is carried out considering all of them 
only partially suggesting a common origin for their anomalous signal related to the low CMB quadrupole.
We have found that the \emph{Planck} constraints on neutrino masses $\sum m_{\nu}$ are essentially stable
under the inclusion of $\betas$. 

We have shown how future measurements of CMB $\mu$-type spectral distortions from the dissipation of acoustic waves, such as those expected by PIXIE, could
severely constrain both the running and the running of the running. More precisely we have found that an improvement on \emph{Planck} bounds by a factor of $\sim 80\%$ is expected. Finally, we discussed the conditions under which the phenomenological expansion of the primordial power spectrum in \eq{Delta_of_k} can be extended to scales much shorter that those probed by CMB anisotropies and can provide a good approximation to the predictions of inflationary models.

%--> We were a bit worried by two things:
%-) first of all, the Taylor series for ns(k) begins to lose validity around \mu-distortion scales, for nrunrun ~ 0.02;
%-) on top of that, for these values of nrun and nrunrun, the extrapolation of the inflationary power spectrum fails much before reaching those scales.
%We think that the first problem (forgetting about inflation) can be addressed by saying this: with PIXIE the bounds are tightened enough that the expansion becomes valid "a posteriori".
%Besides, I thought that it was enough to present our argument by using what you called "Method II" in your paper http://arxiv.org/abs/1603.02496, since the predictions for \mu would vary a lot at these large values of nrun and nrunrun. However, I would like to ask your opinion on this.

\section*{Acknowledgments}

\noindent We would like to thank Jens Chluba and Takeshi Kobayashi for careful reading of the manuscript and useful comments. We would also like to thank the referee for providing useful comments. GC and AM are supported by the research grant Theoretical Astroparticle Physics number 2012CPPYP7 under the program PRIN 2012 funded by MIUR and by TASP, iniziativa specifica INFN. EP is supported by the Delta-ITP consortium, a program of the Netherlands organization for scientific research (NWO) that is funded by the Dutch Ministry of Education, Culture and Science (OCW). This work has been done within the Labex ILP (reference ANR-10-LABX-63) part of the Idex SUPER, and received financial state aid managed by the Agence Nationale de la Recherche, as part of the programme Investissements d'avenir under the reference ANR-11-IDEX-0004-02. EDV acknowledges the support of the European Research Council via the Grant number 267117 (DARK, P.I. Joseph Silk).

\section{Appendix}
\label{sec:appendix}

\subsection{Dependence on the choice of pivot scale}
\label{sec:dependence_pivot}

\begin{figure*}
\begin{center}
\begin{tabular}{c c}
\includegraphics[width=\columnwidth]{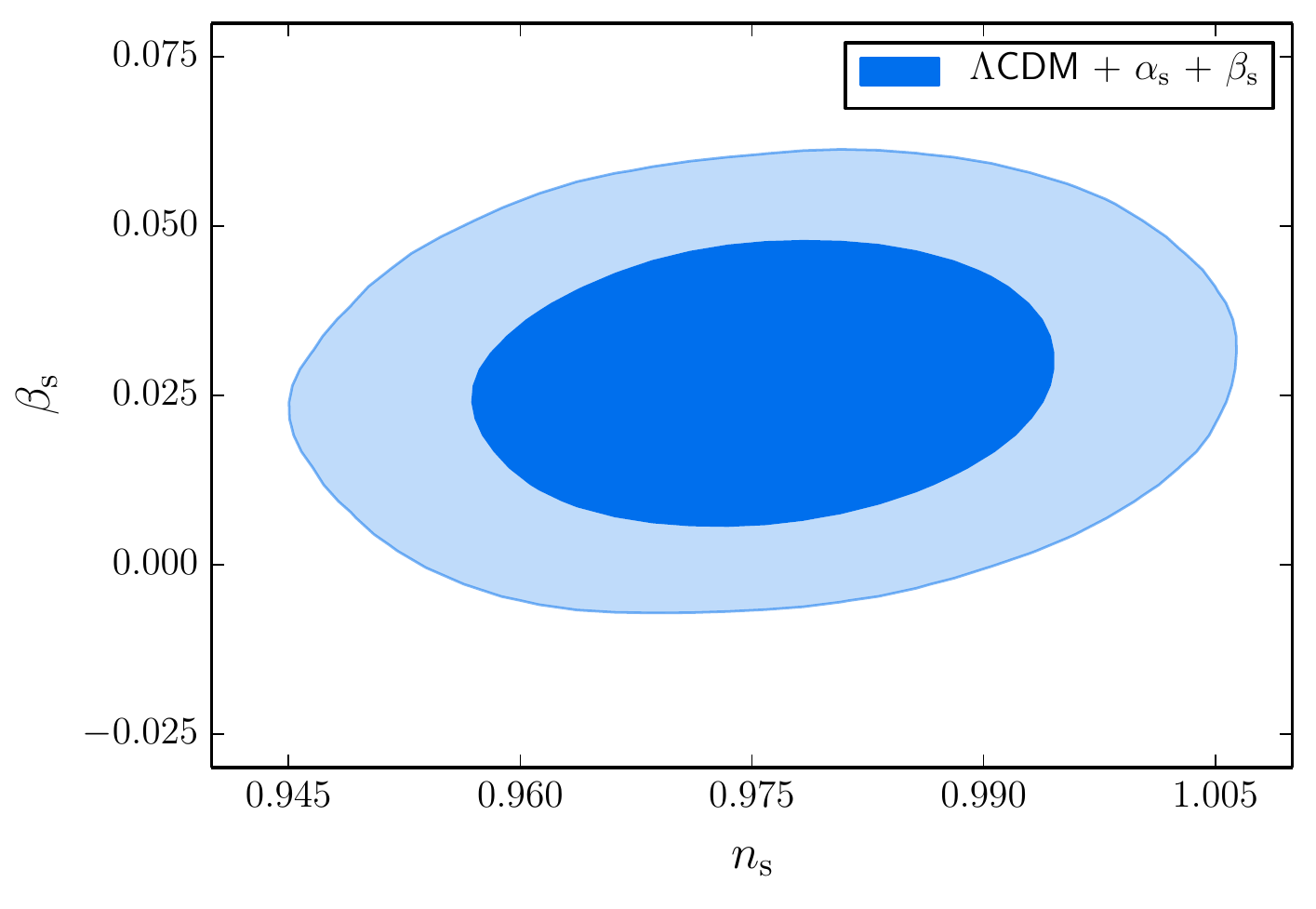}
&\includegraphics[width=\columnwidth]{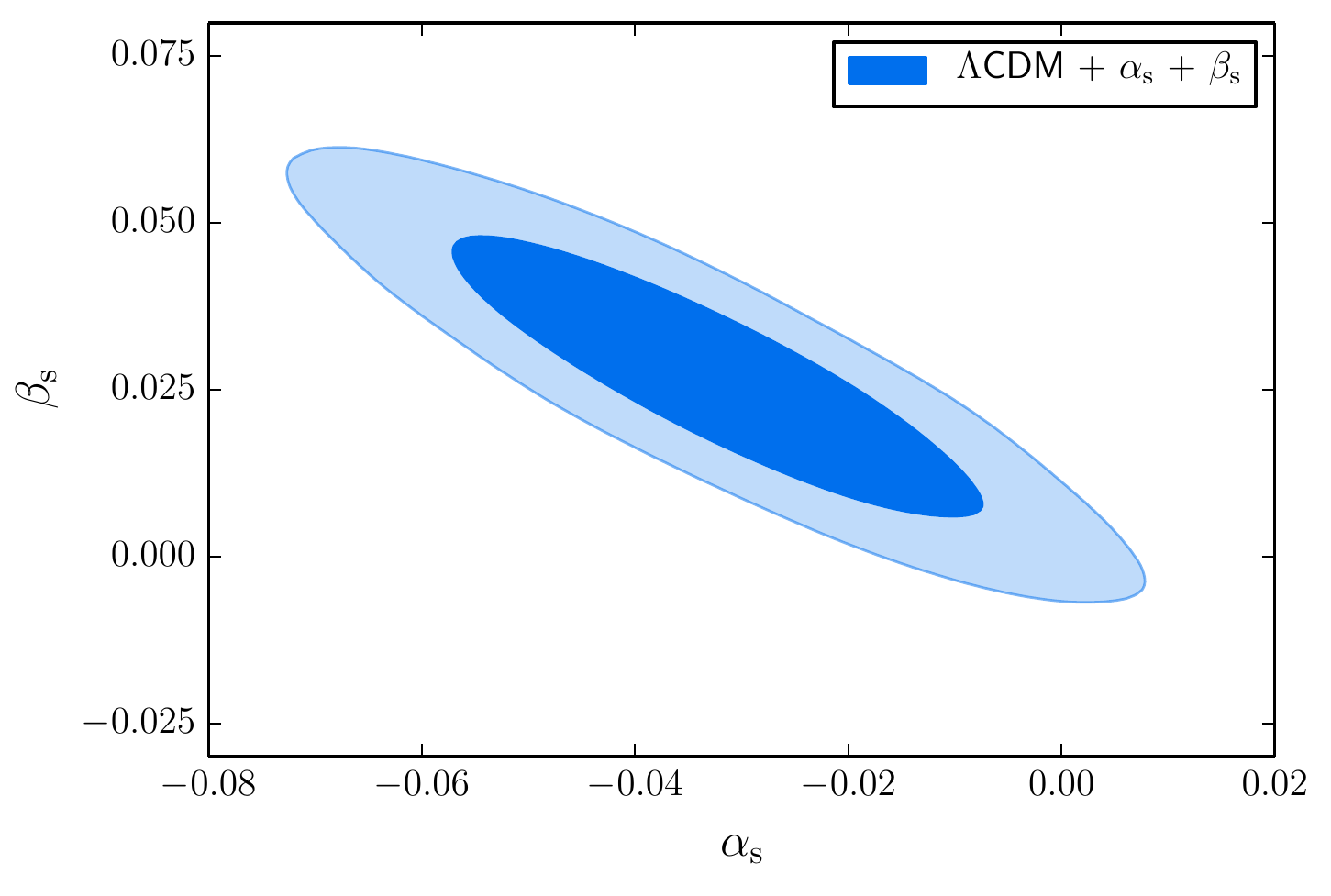}
\end{tabular}
\end{center}
\caption{\footnotesize{Likelihood constraints 
in the $\text{$\ns$ -- $\betas$}$ (left panel) and $\text{$\alphas$ -- $\betas$}$ (right panel) planes for
\emph{Planck} ($TT$, $TE$, $EE$ + lowP), at a pivot $k_\star = 0.01\,\mathrm{Mpc}^{-1}$.}}
\label{fig:pivot-0_01}
\end{figure*}

\begin{figure*}
\begin{center}
\begin{tabular}{c c}
\includegraphics[width=\columnwidth]{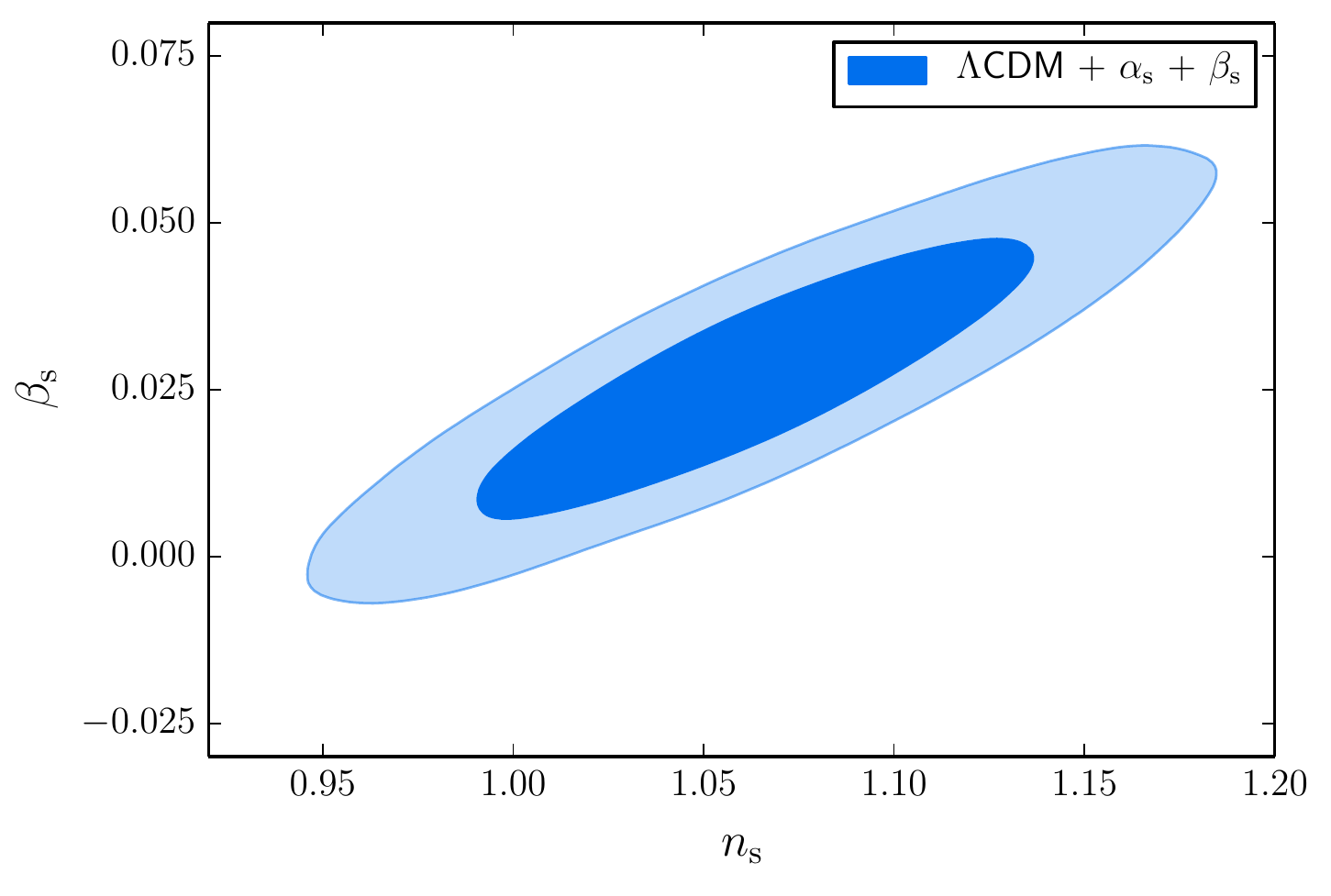}
&\includegraphics[width=\columnwidth]{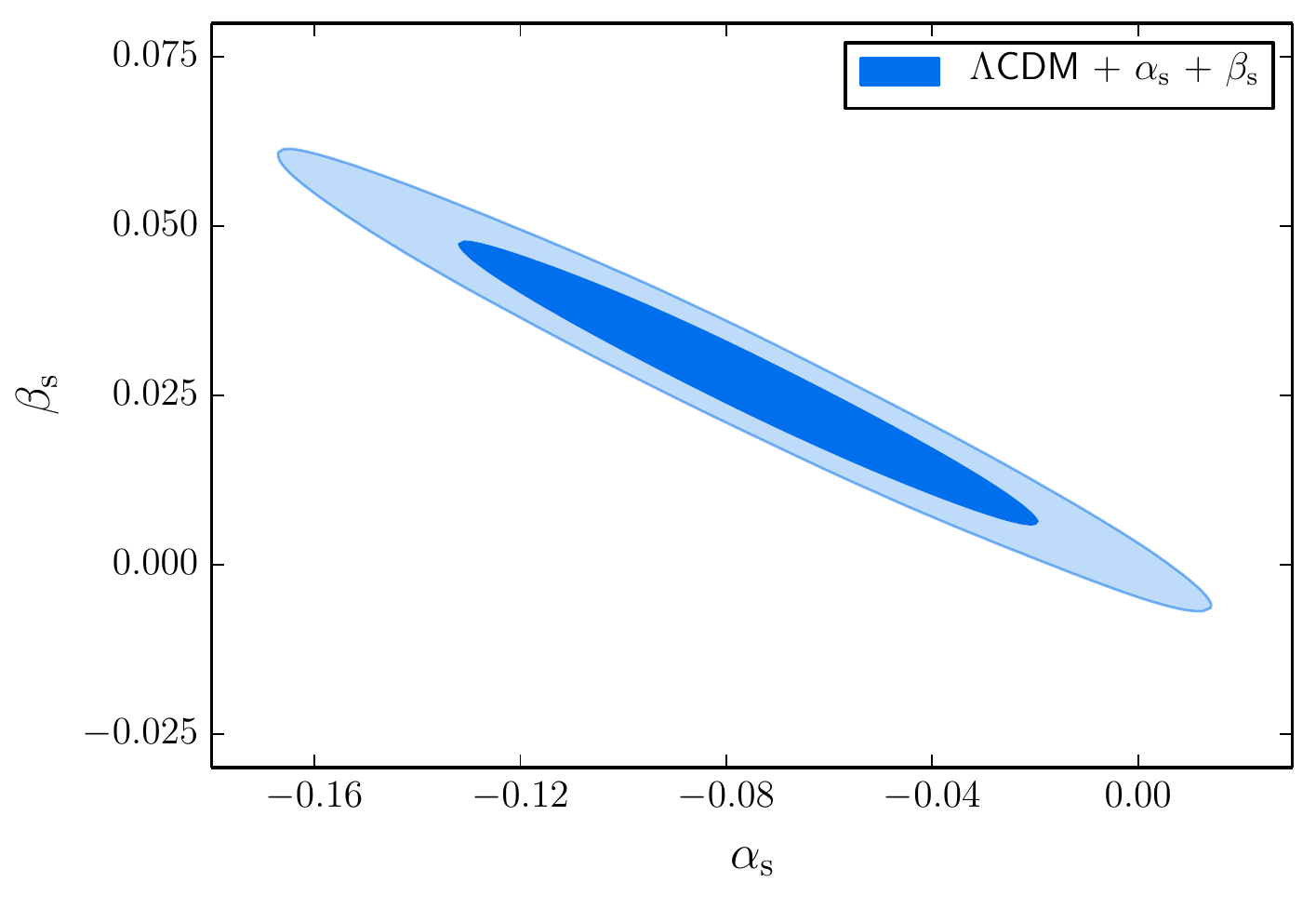}
\end{tabular}
\end{center}
\caption{\footnotesize{Likelihood constraints 
in the $\text{$\ns$ -- $\betas$}$ (left panel) and $\text{$\alphas$ -- $\betas$}$ (right panel) planes for
\emph{Planck} ($TT$, $TE$, $EE$ + lowP), at a pivot $k_\star = 0.002\,\mathrm{Mpc}^{-1}$.}}
\label{fig:pivot-0_002}
\end{figure*}

\noindent \textcolor{black}{When including derivatives of the scalar spectral index as free parameters, one can expect that the constraints on them will depend on the choice of pivot scale $k_\star$ \cite{Cortes:2007ak}: for example, for \emph{Planck} the pivot $k_\star = 0.05\,\mathrm{Mpc}^{-1}$ is chosen to decorrelate $\ns$ and $\alphas$. For this reason, we considered two additional values of $k_\star$ in the analysis of the ``base'' ($\Lambda\mathrm{CDM} + \alphas + \betas$) model with \emph{Planck} ($TT$, $TE$, $EE$ + lowP) data: $k_\star = 0.01\,\mathrm{Mpc}^{-1}$ and $k_\star = 0.002\,\mathrm{Mpc}^{-1}$. We report the results in Figs.~\ref{fig:pivot-0_01}, \ref{fig:pivot-0_002} and \tab{pivots}: we see that at $k_\star = 0.01\,\mathrm{Mpc}^{-1}$ the tilt and $\betas$ decorrelate, while the degeneracy between $\alphas$ and $\betas$ goes from positive to negative. For $k_\star = 0.002\,\mathrm{Mpc}^{-1}$, instead, we see that $\alphas$ and $\betas$ are still negatively correlated, while the degeneracy between $\ns$ and $\betas$ becomes positive. However we see from \tab{pivots} that, while changing the pivot cancels the $1\sigma$ indication for $\alphas > 0$, the $2\sigma$ preference for $\betas > 0$ remains in both cases.}

\textcolor{black}{We can understand why the marginalized error on $\betas$ does not change if we change the pivot scale $k_\star$ with a simple Fisher analysis. For a log-likelihood for $\vec{n}\equiv(\ns,\alphas,\betas)$ (marginalized over all parameters except $\ns$, $\alphas$, $\betas$) given by} 
\begin{equation}
\label{eq:likelihood_n}
\mathcal{L}|_{k_\star^{(0)}}\propto (\vec{n}-\vec{n}_0)^T\cdot F_{k_\star^{(0)}}\cdot(\vec{n}-\vec{n}_0)\,\,,
\end{equation}
\textcolor{black}{with inverse covariance matrix $F_{k_\star^{(0)}}$, a change of pivot will result in} 
\begin{equation}
\label{eq:likelihood_n_transformed}
\mathcal{L}|_{k_\star}\propto (M\cdot\vec{n}-\vec{n}_0)^T\cdot F_{k_\star^{(0)}}\cdot(M\cdot\vec{n}-\vec{n}_0)\,\,,
\end{equation}
\textcolor{black}{where $M$ is given by the scale dependence of $\vec{n}$, \emph{i.e.}}
\begin{equation}
\label{eq:M_matrix}
\begin{split}
\vec{n}_{k_\star} &= M\cdot\vec{n}_{k_\star^{(0)}} \\
&= 
\begin{pmatrix}
1 & \log\frac{k_\star}{k_\star^{(0)}} & \frac{1}{2}\log^2\frac{k_\star}{k_\star^{(0)}}\\
0 & 1 & \log\frac{k_\star}{k_\star^{(0)}} \\
0 & 0 & 1
\end{pmatrix}
\begin{pmatrix}
\ns(k_\star^{(0)}) \\
\alphas(k_\star^{(0)}) \\
\betas(k_\star^{(0)})
\end{pmatrix}
\,\,,
\end{split}
\end{equation}
\textcolor{black}{and it is straightforward to verify that it has unit determinant. For a Gaussian likelihood, we can forget about $\vec{n}_0$ (we can just call $\vec{n}_0 = M\cdot\vec{m}_0$ and do a translation), so that all information will be coming from the transformed inverse covariance, \emph{i.e.}}
\begin{equation}
\label{eq:transformed_F}
F_{k_\star} = M^T\cdot F_{k_\star^{(0)}}\cdot M\,\,.
\end{equation}
 
\textcolor{black}{Since $M$ has unit determinant, the ``figure of merit'' $\text{f.o.m.}\propto 1/\det{F_{k_\star}}$ (which is basically $1/\text{volume of $\limit{68}$ ellipsoid}$) will not change if we change the pivot. What will indeed change are the marginalized and non-marginalized $1\sigma$ errors on the parameters: however, it is straightforward to show with linear algebra that the marginalized error on the running of the running, which is given by} 
\begin{equation}
\label{eq:one_sigma_beta}
\sigma(\betas(k_\star)) = \sqrt{\Big(F_{k_\star}^{-1}\Big)_{33}}\,\,,
\end{equation}
\textcolor{black}{does not change under the transformation of \eq{M_matrix}.}

\textcolor{black}{This simple picture does not explain why the mean values of $\ns$ and $\alphas$ change. We ascribe this to the presence of the additional parameter $A_\mathrm{s}$: under the transformation of \eq{M_matrix} it will not change linearly, so the Gaussian approximation will not hold. The data will still constrain $A_\mathrm{s}$ well enough, so that $\sigma(A_\mathrm{s})$ will not contribute to the errors on the parameters, but the position of the peak of the transformed likelihood will change.}

\subsection{$\Delta\chi^2$: base model vs. extensions}
\label{sec:chi2}

\noindent\textcolor{black}{In this appendix we collect the full $\Delta\chi^2$ tables: we refer to \sect{results} for a discussion of the various improvements and non-improvements in $\chi^2$ for the different choices of datasets and parameters that have been considered. In all the tables below, $\Delta\chi^2$ stands for $\chi^2_\mathrm{base} - \chi^2_\text{base + ext.}$, both obtained via MCMC sampling of the likelihood.}

\begin{table}[!hbtp]
\begin{center}
\begin{tabular}{lcccc}
\toprule
\horsp
 \vertsp vs. ${} + A_L$ \vertsp vs. ${} + \sum m_\nu$ \vertsp vs. ${} + \Omega_K$ \\
\hline
\morehorsp
$\Delta\chi^2_\mathrm{plik}$ \vertsp \siround{2.1}{1} \vertsp \siround{-1.8}{1} \vertsp \siround{2.4}{1} \\
\morehorsp
$\Delta\chi^2_\mathrm{lowP}$ \vertsp \siround{-0.9}{1} \vertsp \siround{-0.6}{1} \vertsp \siround{-1.3}{1} \\
\morehorsp
$\Delta\chi^2_\mathrm{prior}$ \vertsp \siround{-1.}{1} \vertsp \siround{0.1}{1} \vertsp \siround{-1.9}{1} \\
\hline
\morehorsp
$\Delta\chi^2$ \vertsp \siround{0.2}{1} \vertsp \siround{-2.3}{1} \vertsp \siround{-0.7}{1} \\
\hline
\bottomrule
%\botrule
\end{tabular}
\caption{\footnotesize{$\chi^2$ comparison between the base $\Lambda\mathrm{CDM} + \alphas + \betas$ model and the other extensions considered in the main text, for the \emph{Planck} $TT$, $TE$, $EE$ + lowP dataset. The last line contains the overall $\Delta\chi^2$ for all the likelihoods included in the analysis.}}
\label{tab:chi2-pl}
\end{center}
\end{table}

\begin{table}[!hbtp]
\begin{center}
\begin{tabular}{lcccc}
\toprule
\horsp
 \vertsp vs. ${} + A_L$ \vertsp vs. ${} + \sum m_\nu$ \vertsp vs. ${} + \Omega_K$ \\
\hline
\morehorsp
$\Delta\chi^2_\mathrm{plik}$ \vertsp \siround{1.9}{1} \vertsp \siround{-0.5}{1} \vertsp \siround{1.5}{1} \\
\morehorsp
$\Delta\chi^2_\mathrm{lowP}$ \vertsp \siround{-0.5}{1} \vertsp \siround{0.1}{1} \vertsp \siround{0.}{1} \\
\morehorsp
$\Delta\chi^2_\mathrm{prior}$ \vertsp \siround{-3.8}{1} \vertsp \siround{-2.7}{1} \vertsp \siround{-0.1}{1} \\
\morehorsp
$\Delta\chi^2_\mathrm{lensing}$ \vertsp \siround{0.9}{1} \vertsp \siround{1.5}{1} \vertsp \siround{-1.3}{1} \\
\hline
\morehorsp
$\Delta\chi^2$ \vertsp \siround{-1.6}{1} \vertsp \siround{-1.6}{1} \vertsp \siround{0.1}{1} \\
\hline
\bottomrule
%\botrule
\end{tabular}
\caption{\footnotesize{Same as \tab{chi2-pl}, but with the addition of CMB lensing data.}}
\label{tab:chi2-pl+lens}
\end{center}
\end{table}

\begin{table}[!hbtp]
\begin{center}
\begin{tabular}{lcccc}
\toprule
\horsp
 \vertsp vs. ${} + A_L$ \vertsp vs. ${} + \sum m_\nu$ \vertsp vs. ${} + \Omega_K$ \\
\hline
\morehorsp
$\Delta\chi^2_\mathrm{plik}$ \vertsp \siround{3.}{1} \vertsp \siround{-4.3}{1} \vertsp \siround{-1.6}{1} \\
\morehorsp
$\Delta\chi^2_\mathrm{lowP}$ \vertsp \siround{-0.3}{1} \vertsp \siround{0.8}{1} \vertsp \siround{-0.3}{1} \\
\morehorsp
$\Delta\chi^2_\mathrm{prior}$ \vertsp \siround{0.8}{1} \vertsp \siround{2.9}{1} \vertsp \siround{0.1}{1} \\
\morehorsp
$\Delta\chi^2_\mathrm{CFHTLenS}$ \vertsp \siround{2.3}{1} \vertsp \siround{-0.6}{1} \vertsp \siround{3.8}{1} \\
\hline
\morehorsp
$\Delta\chi^2$ \vertsp \siround{5.9}{1} \vertsp \siround{-1.3}{1} \vertsp \siround{2.}{1} \\
\hline
\bottomrule
%\botrule
\end{tabular}
\caption{\footnotesize{Same as \tab{chi2-pl}: the dataset is \emph{Planck} $TT$, $TE$, $EE$ + lowP + WL.}}
\label{tab:chi2-pl+wl}
\end{center}
\end{table}

\begin{table}[!hbtp]
\begin{center}
\begin{tabular}{lcccc}
\toprule
\horsp
 \vertsp vs. ${} + A_L$ \vertsp vs. ${} + \sum m_\nu$ \vertsp vs. ${} + \Omega_K$ \\
\hline
\morehorsp
$\Delta\chi^2_\mathrm{plik}$ \vertsp \siround{0.7}{1} \vertsp \siround{1.}{1} \vertsp \siround{-2.7}{1} \\
\morehorsp
$\Delta\chi^2_\mathrm{lowP}$ \vertsp \siround{-1.8}{1} \vertsp \siround{-1.2}{1} \vertsp \siround{-1.}{1} \\
\morehorsp
$\Delta\chi^2_\mathrm{prior}$ \vertsp \siround{1.4}{1} \vertsp \siround{0.3}{1} \vertsp \siround{0.8}{1} \\
\morehorsp
$\Delta\chi^2_\mathrm{6DF}$ \vertsp \siround{0.1}{1} \vertsp \siround{0.}{1} \vertsp \siround{0.1}{1} \\
\morehorsp
$\Delta\chi^2_\mathrm{MGS}$ \vertsp \siround{-0.8}{1} \vertsp \siround{0.}{1} \vertsp \siround{-0.8}{1} \\
\morehorsp
$\Delta\chi^2_\mathrm{DR11CMASS}$ \vertsp \siround{0.9}{1} \vertsp \siround{0.1}{1} \vertsp \siround{1.1}{1} \\
\morehorsp
$\Delta\chi^2_\mathrm{DR11LOWZ}$ \vertsp \siround{1.1}{1} \vertsp \siround{0.1}{1} \vertsp \siround{1.1}{1} \\
\hline
\morehorsp
$\Delta\chi^2$ \vertsp \siround{1.5}{1} \vertsp \siround{0.3}{1} \vertsp \siround{-1.4}{1} \\
\hline
\bottomrule
%\botrule
\end{tabular}
\caption{\footnotesize{$\Delta\chi^2$ for the \emph{Planck} $TT$, $TE$, $EE$ + lowP + BAO dataset.}}
\label{tab:chi2-pl+bao}
\end{center}
\end{table}

\subsection{%Slow-roll, $\alphas$ and $\betas$
Derivation of slow-roll expansion for $\epsilon$}
\label{sec:appendix-slow_roll}

\noindent Starting from \eq{slow_roll_rels-2}, differentiating it w.r.t. $N$ and then using \eq{slow_roll_rels-1}, one can find the coefficients $\epsilon^{(i)}$ of a Taylor expansion of $\epsilon(N)$ in terms of the parameters describing the scale dependence of the primordial spectrum $\Delta^2_\zeta(k)$. More precisely, one finds (calling $\epsilon_\star\equiv\epsilon(N_\star)$)
\begin{subequations}
\label{eq:epsilon_coefficients}
\begin{align}
&\epsilon^{(1)} = (\ns - 1)\epsilon_\star + 2\epsilon_\star^2\,\,, \label{eq:epsilon_coefficients-1} \\
&\epsilon^{(2)} = -\alphas\epsilon_\star + 4\epsilon_\star\epsilon^{(1)} + (\ns-1)\epsilon^{(1)}\,\,, \label{eq:epsilon_coefficients-2} \\
&\epsilon^{(3)} = \betas\epsilon_\star - 2\alphas\epsilon^{(1)} \nonumber \\
&\hphantom{\epsilon^{(3)} =} + (\ns-1)\{-\alphas\epsilon_\star + 4\epsilon_\star\epsilon^{(1)} + (\ns-1) \epsilon^{(1)}\} \nonumber \\
&\hphantom{\epsilon^{(3)} =} + 4\{\epsilon_\star[-\alphas\epsilon_\star + 4\epsilon_\star\epsilon^{(1)}+(\ns-1)\epsilon^{(1)}] + (\epsilon^{(1)})^2\}\,\,. \label{eq:epsilon_coefficients-3}
\end{align}
\end{subequations}
By plugging in the values of $\alphas$ and $\betas$ allowed by \emph{Planck}, one can extrapolate $\epsilon$ at scales different from $k_\star$. See \sect{slow_roll_inflation} for a discussion.

%%%%%%%%%%%%%%%%%%%%%%%%%%%%%%%%%%%%%%%%%%%%%%%%%%%%%%%%%%%%%%%%%%%%%%%%%%%%%%%%%

\end{document}